\documentclass[aps,pra,twocolumn,amssymb,amsmath]{revtex4-2}
\usepackage{graphicx}
\usepackage{bm}

\begin{document}
\title{Axiomatic foundation of quantum measurements and survival effect}

\author{Vladimir I. Kruglov}

\affiliation{Centre for Engineering Quantum Systems, 
School of Mathematics and Physics, The University of Queensland, Brisbane, QLD 4072, Australia}

\begin{abstract}
The axiomatic theory of quantum first-kind measurements is developed in a rigorous form based on five Postulates. The measurement theory for observable with continuous spectrum is given in a rigged Hilbert space. This approach also describes the measurements with non-ideal initial conditions. It yields the survival effect in the position measurement of the particles. It is also found that there is no such survival effect in the momentum measurement of the particles. These Postulates of axiomatic theory yield 
the survival effect which violates the Heisenberg uncertainty relation. This theoretical result is demonstrated by the wave function with minimum of position and momentum uncertainty of the particle. The survival effect leads to essential corrections for the uncertainty relations. These modified uncertainty relations can also be used for the experimental verification of the survival measurement effect.

\end{abstract}


\maketitle

\section{Introduction}

It is well known that the measuring process plays an important  role in quantum mechanics \cite{P}. The first formal model
of such process was developed by von Neumann \cite{JN}, however his point of view has been criticized (see for an example \cite{WIG, Lud,AY,Wak,DLP}). Many other authors have made attempts to obtain a more satisfactory solution of the quantum measurement problem based on the principles of quantum mechanics and statistical physics. In more recent papers, authors propose the particular schemes and models of quantum measurement with unitary and non-unitary evolution \cite{CM,C,SCW,WM,WM2,WM3,WM4,WM5,MB,RLBW,RLW,RL,WIS,Mil,X}. The evolution of measuring quantum system is  modelling by stochastic equations in Refs. \cite{DZR,DPZ,GPZ,TM,VS,CSR}. 

In the traditional formulation of quantum mechanics in
Hilbert space, states are described as density operators
and observables are represented as self-adjoint operators.
Alternatively, and equivalently, experimental events and
propositions are represented as orthogonal projection operators, and states are defined as generalized probability measures. The axiomatic formulation of quantum measurements presented in this paper shows that the satisfactory solution of the quantum measurement problem can be based on a rigorous approach using the rigged Hilbert space, defined in Appendix B, and appropriate set of Postulates describing the measurement process. The Postulates describe, or define, in this approach the quantum nature of measuring devices and their quantum evolution based on general scattering theory given in Appendix A.

In this paper, we present the axiomatic theory of quantum first-kind measurements in rigorous form. This approach also allows us to declare the theory of non-ideal measurements. Thus, there are two main purposes of this study. First of all we show that the general quantum scattering theory is consistent with
the quantum measurements formulated in Quantum Mechanics. Secondly, using this rigorous approach we
study the survival effect which leads to essential corrections for the uncertainty relations. Moreover,
these modified uncertainty relations can also be used for the experimental verification of the survival measurement effect.

In Section II, the main set of postulates and the basic definitions of the axiomatic theory is formulated. In Sections III and IV, we consider the measurements of an observable with discrete spectrum, and in Section V, it is presented the description of momentum and position of the particle measurement using a rigged Hilbert space. In Section VI, a non-ideal measurement theory is developed. This approach leads to survival effect for a discrete and continuous spectrum measurements. It is shown that no such effect in the momentum measurement of the particles. In Section VII, we study the survival effect for the position measurement of the particles. 
It is shown that the survival effect leads to modified (weak) form for the Heisenberg uncertainty relation. This effect is demonstrated by the wave function with minimum of position and momentum uncertainty of the particle.
In Sections VIII and IX, we consider in more details different aspects of the survival effect leading to modified form for the Heisenberg uncertainty relation and appropriate modification of other uncertainty relations. We emphasize that these modified uncertainty relations have a simple physical interpretation.

\section{Axiomatics of quantum measurement theory}

In this section, we formulate the set of postulates of the axiomatic theory of quantum first-kind measurements.
These postulates are formulated for the case when the measured observable has a discrete degenerate or non-degenerate spectrum. The postulates and definitions formulated in this section are universal because an observable with continuous spectrum has in the rigged Hilbert space (RHS) a discrete representation (see Sec. V).   

 Let ${\cal A}$ be an observable which has a discrete degenerate spectrum with eigenvalues $\lambda_\alpha$ and eigenstates $|u_\alpha^s\rangle$, 
\begin{equation}
{\cal A}|u_\alpha^s\rangle=\lambda_\alpha|u_\alpha^s\rangle,~~~\sum_\alpha \sum_{s=1}^{g_\alpha} |u_\alpha^s\rangle \langle u_\alpha^s|={\bf 1}_{\rm Q},
\label{1}
\end{equation}
where $g_\alpha$ is the degeneracy of the state $|u_\alpha^s\rangle$ and the eigenstates are normalized $\langle u_\alpha^n|u_\beta^s\rangle=\delta_{\alpha\beta}\delta_{ns}$. Here ${\bf 1}_{\rm Q}$ is the unit operator in the space of measured quantum system. Note that the case when an observable has a discrete degenerate spectrum is a particular case of an observable with non-degenerate spectrum. In general case the observable ${\cal A}$ can be written in the form,
\begin{equation}
{\cal A}=\sum_\alpha \lambda_\alpha {\cal P}_\alpha,~~~{\cal P}_\alpha=\sum_{s=1}^{g_\alpha} |u_\alpha^s\rangle \langle u_\alpha^s|,
\label{2}
\end{equation}
where ${\cal P}_\alpha$ is the projector into the subspace ${\cal G}_\alpha$ of ${\cal A}$.
The eigenstates in Eq. (\ref{1}) and the projector ${\cal P}_\alpha$ can also be written in a new orthonormal basis,
\begin{equation}
{\cal P}_\alpha=\sum_{s=1}^{g_\alpha} |\tilde{u}_\alpha^s\rangle \langle \tilde{u}_\alpha^s|,~~~\langle \tilde{u}_\alpha^n|\tilde{u}_\beta^s\rangle=\delta_{\alpha\beta}\delta_{ns}.
\label{3}
\end{equation}
The states $|u_\alpha^s\rangle$ and $|\tilde{u}_\alpha^s\rangle$ are connected by unitarian matrix $\Lambda_\alpha$, 
\begin{equation}
|\tilde{u}_\alpha^s\rangle=\sum_{n=1}^{g_\alpha}\Lambda_\alpha^{sn}|u_\alpha^n\rangle,~~~\Lambda_\alpha^{\dag} \Lambda_\alpha=\Lambda_\alpha \Lambda_\alpha^{\dag}={\bf 1}_\alpha,
\label{4}
\end{equation}
where ${\bf 1}_\alpha$ is a unit matrix in the subspace ${\cal G}_\alpha$. Eq. (\ref{1}) in a new orthonormal basis has the form, 
\begin{equation}
{\cal A}|\tilde{u}_\alpha^s\rangle=\lambda_\alpha|\tilde{u}_\alpha^s\rangle,~~~\sum_\alpha \sum_{s=1}^{g_\alpha} |\tilde{u}_\alpha^s\rangle \langle \tilde{u}_\alpha^s|={\bf 1}_{\rm Q},
\label{5}
\end{equation}

Let ${\cal G}_Q$ is the space of measurable quantum system (QS) and ${\cal G}_D={\cal G}_M\otimes{\cal G}_E$ is the space
of measuring device (MD) interacting with an environment system (ES).
Here ${\cal G}_M$ is the space of measuring device and ${\cal G}_E$ is the space of an environment system. Thus ${\cal G}_Q\otimes{\cal G}_D={\cal G}_Q\otimes{\cal G}_M\otimes{\cal G}_E$ is the space of combined quantum system QS+MD+ES. The measurable QS is described by the density matrix,
\begin{equation}
\rho(t)=\sum_k \pi_k |\psi_k(t)\rangle \langle \psi_k(t)|,
\label{6}
\end{equation}
where $\sum_k \pi_k=1$ ($\pi_k>0$) and $|\psi_k(t)\rangle$ is normalized ket vector $\langle\psi_k(t)|\psi_k(t)\rangle=1$. In general case the initial metastable state of the system MD+ES is given by density matrix $\rho^D_{in}(t)$ in a space ${\cal G}_D={\cal G}_M\otimes{\cal G}_E$ as
\begin{equation}
\rho^D_{in}(t)=\int |\Phi_{in}(Z,t)\rangle {\rm Q}(Z,Z')\langle \Phi_{in}(Z',t)| d\mu(Z)d\mu(Z'),
\label{7}
\end{equation}
where $\langle \Phi_{in}(Z,t)|\Phi_{in}(Z,t)\rangle=1$ and ${\rm Tr}\rho^D_{in}(t)=1$. Here $|\Phi_{in}(Z,t)\rangle$ is the ket vector describing some state of combined quantum system MD+ES and $\mu(Z)$ is the measure in the space ${\cal G}_D$. We emphasize that the density matrix $\rho^D_{in}(t)$
describes the metastable initial state of the system MD+ES. 
Thus, the states of combined system QS+MD+ES are defined in a space ${\cal G}_Q\otimes{\cal G}_M\otimes{\cal G}_E$.  The initial metastable state of measuring device interacting with ES is given by density matrix $\rho^D_{in}(t)$ in Eq. (\ref{7}) and the density matrix $\rho^D_{in}(t)$ is defined in a space ${\cal G}_D={\cal G}_M\otimes{\cal G}_E$. 

{\it Postulate 1: 
The possible results of the measurement of an observable ${\cal A}$ with a discrete degenerate spectrum or non-degenerate spectrum are the eigenvalues $\lambda_\alpha$ of this observable. The measuring device after the measurement of observable ${\cal A}$ indicates some particular eigenvalue $\lambda_\alpha$}. 
\medskip

Note that one can consider this Postulate as a definition of measuring device. 
The full Hamiltonian of the combined quantum system QS+MD+ES can be written in the form, 
\begin{equation}
{\rm H}={\rm H}_0+{\rm H}_{\rm I},~~~{\rm H}_0={\rm H}_{\rm Q}\otimes{\bf 1}_{\rm D}+{\bf 1}_{\rm Q}\otimes{\rm H}_{\rm D}.
\label{8}
\end{equation}
The Hamiltonian ${\rm H}_0$ describes a free evolution of two systems: QS by the Hamiltonian ${\rm H}_{\rm Q}$ in the space ${\cal G}_Q$ and MD+ES by the Hamiltonian ${\rm H}_{\rm D}$ in the space ${\cal G}_M\otimes{\cal G}_E$. Note that the Hamiltonian ${\rm H}_{\rm D}$ includes the interaction between 
MD and ES. The Hamiltonian ${\rm H}_{\rm I}$ describing the interaction of QS and the system MD+ES is defined in the space ${\cal G}_Q\otimes{\cal G}_M\otimes{\cal G}_E$. 
 
 The solution of the Schr\"{o}dinger equation with the Hamiltonian ${\rm H}$ and the boundary conditions for incoming waves at $t<0$
 is given by $|\Psi(t)\rangle={\cal U}_\nu^{in}(t)|\Phi\rangle$ where ${\cal U}_\nu^{in}(t)={\rm U}(t)\Omega_\nu^{(+)}$ is
 the scattering wave propagator. This propagator is defined in Eq. (\ref{3a}) and the wave operator $\Omega_\nu^{(+)}$ is given in Eq. (\ref{4a}) of Appendix A. We can also define the scattering wave propagator ${\cal U}_\nu^{out}(t)$ which yields the formal solution 
 $|\Psi(t)\rangle={\cal U}_\nu^{out}(t)|\Phi\rangle$ of the Schr\"{o}dinger equation with the Hamiltonian ${\rm H}_0$ and the boundary conditions for outcoming waves at $t>0$. The scattering wave propagator ${\cal U}_\nu^{out}(t)$ defined with such boundary conditions is given by
\begin{eqnarray}
{\cal U}_\nu^{out}(t)=\nu\int_0^{+\infty} dt' e^{-\nu t'}{\rm U}_0(t-t'){\rm U}(t')~~~~~~~~~
\nonumber\\ \noalign{\vskip3pt}
={\rm U}_0(t)\nu\int_0^{+\infty} dt' e^{-\nu t'}{\rm U}_0^{\dag}(t'){\rm U}(t')={\rm U}_0(t)\Omega_\nu^{(-)\dag},~~~~~~~~~
\label{9}
\end{eqnarray}
where ${\rm U}(t)=\exp(-it{\rm H}/\hbar)$, ${\rm U}_0(t)=\exp(-it{\rm H}_0/\hbar)$ and $\Omega_\nu^{(-)}$ is the wave operator given in Eq. (\ref{5a}).
It is assumed in this definition that $\nu\rightarrow 0^{+}$. Note that this limit can be applied only in the final step for evaluation of the functionals such as probabilities and cross sections. More precise limiting process is presented in the end of Appendix A. 

It is assumed in the paper that the hierarchy of times has the form: $t\geq t_b\geq t_a$ where $t_a$ is the initial point of measuring process, $t_b$ is the final point of measuring process and $t$ is an arbitrary time after the measurement of the observable ${\cal A}$. 
We define the conditional wave propagator (CWP) as 
${\cal U}_\nu(t',t_0)={\cal U}_\nu^{out}(t-t_b){\cal U}_\nu^{in}(t_b-t_a)$ where $t'=t-t_b$ and $t_0=t_b-t_a$. This propagator describes the evolution of states in the system QS+MD+ES as long as the initial point $t_a$ and the final point $t_b$ of measuring
process are fixed. Note that the final point $t_b$ in measuring process is random and hence the CWP describes the conditional outgoing  states.
The conditional wave propagator ${\cal U}_\nu(t',t_0)$ has the form,
\begin{equation}
{\cal U}_\nu(t',t_0)={\rm U}_0(t')\Omega_\nu^{(-)\dag}{\rm U}(t_0)\Omega_\nu^{(+)}={\rm U}_0(t-t_b){\cal K}_\nu(t_0),
\label{10}
\end{equation}
where ${\cal K}_\nu(t_0)=\Omega_\nu^{(-)\dag}{\rm U}(t_0)\Omega_\nu^{(+)}$. 
The operator ${\cal K}_\nu(t_0)$ yields the relations ${\rm lim}_{t_0\rightarrow 0}{\cal K}_\nu(t_0)=S$ and ${\rm lim}_{t_0\rightarrow 0}{\cal U}_\nu(t',t_0)={\rm U}_0(t')S$ where $S=\Omega_\nu^{(-)\dag}\Omega_\nu^{(+)}$ is the scattering operator defined in Eq. (\ref{7a}). In the case when no interaction in the system
 (${\rm H}_{\rm I}=0$) the following relation is valid $\Omega_\nu^{(\pm)}={\bf 1}$. Thus, in this case we have ${\cal K}_\nu(t_0)={\rm U}_0(t_0)$ and ${\cal U}_\nu(t',t_0)={\rm U}_0(t-t_a)$. 
 
The operator describing free evolution (${\rm H}_{\rm I}=0$) is given by ${\rm U}_0(t)={\rm U}_{\rm Q}(t)\otimes{\rm U}_{\rm D}(t)$ where ${\rm U}_{\rm Q}(t)=\exp(-it{\rm H}_{\rm Q}/\hbar)$ and ${\rm U}_{\rm D}(t)=\exp(-it{\rm H}_{\rm D}/\hbar)$. 
 Note that we consider the first-kind measurements. We also suppose that in this case the measurable QS and measuring device do not have the bound states for the interaction Hamiltonian ${\rm H}_{\rm I}$ and hence the wave operators $\Omega_\nu^{(\pm)}$ and propagator ${\cal U}_\nu(t',t_0)$ are unitarian (see Appendix A). The next notations are used below: $F(t',t_0,t_a)\equiv F(T)$ and $F(0,t_0,t_a)\equiv F(T_0)$ where $F$ is an arbitrary function of variables $t', ~t_0,~ t_a$. We also define the eigenstates of non-degenerate observable ${\cal A}$ by relation ${\cal A}|u_\alpha\rangle=\lambda_\alpha|u_\alpha\rangle$ where it is assumed the normalization $\langle u_\alpha|u_\beta\rangle=\delta_{\alpha\beta}$.

{\it Postulate 2: 
The conditional evolution of the initial state $|u_\alpha\rangle \otimes |\Phi_{in}(Z)\rangle$ 
of combined system QS+MD+ES with fixed initial point $t_a$ and final point $t_b$ 
is given by
\begin{equation}
{\cal U}_\nu(t',t_0)|u_\alpha\rangle \otimes |\Phi_{in}(Z,t_a)\rangle = |u_\alpha(t')\rangle \otimes |\Phi_{\alpha}(Z,T)\rangle.
\label{11}
\end{equation}
 The conditional evolution of the initial state $|u_\alpha^s\rangle \otimes |\Phi_{in}(Z)\rangle$ of combined system QS+MD+ES with fixed initial point $t_a$ and final point $t_b$ has the form,
\begin{equation}
{\cal U}_\nu(t',t_0)|u_\alpha^s\rangle \otimes |\Phi_{in}(Z,t_a)\rangle =|\tilde{u}_\alpha^s(t')\rangle \otimes|\Phi_{\alpha}(Z,T)\rangle.
\label{12}
\end{equation}}
\medskip

We note that in Eq. (\ref{11}) we have $|u_\alpha(t')\rangle  ={\rm U}_{\rm Q}(t')|u_\alpha\rangle$ which is a free evolution of the eigenstate of non-degenerate observable
and $|\Phi_{\alpha}(Z,T)\rangle={\rm U}_{\rm D}(t')|\Phi_{\alpha}(Z,T_0)\rangle$. The quantum state $|\Phi_{\alpha}(Z,T)\rangle$ describes the system MD+ES in the space ${\cal G}_D={\cal G}_M\otimes{\cal G}_E$ and the conditional wave propagator ${\cal U}_\nu(t',t_0)$ is given in Eq. (\ref{10}). We have defined
in Eq. (\ref{12}) the state $|\tilde{u}_\alpha^s(t')\rangle ={\rm U}_{\rm Q}(t')|\tilde{u}_\alpha^s\rangle$ which is a free evolution of the eigenstate of degenerate observable and $|\Phi_{\alpha}(Z,T)\rangle={\rm U}_{\rm D}(t')|\Phi_{\alpha}(Z,T_0)\rangle$. The quantum state $|\tilde{u}_\alpha^s\rangle$ is given in Eq. (\ref{4}) where $\Lambda_\alpha$ is some unitarian matrix.

It follows from Eqs. (\ref{11}) and (\ref{12}) that
the quantum state $|\Phi_{\alpha}(Z,T)\rangle$ in Eqs. (\ref{11}) and (\ref{12}) are normalized $\langle\Phi_{\alpha}(Z,T)|\Phi_{\alpha}(Z,T)\rangle=1$. In the case when the density matrix of QS at $t=t_a$ is $\rho_\alpha=|u_\alpha\rangle  \langle u_\alpha|$ then Eq. (\ref{11}) yields 
\begin{equation}
{\cal U}_\nu(t',t_0)[\rho_\alpha \otimes \rho^D_{in}(t_a)]{\cal U}_\nu^\dag(t',t_0)=|u_\alpha(t')\rangle  \langle u_\alpha(t')| \otimes \rho^D_{\alpha}(T).
\label{13}
\end{equation}
If ${\cal A}$ is the degenerate observable and the density matrix of QS at $t=t_a$
is $\rho_\alpha^s=|u_\alpha^s\rangle  \langle u_\alpha^s|$ then Eq. (\ref{12}) yields the relation,
\begin{equation}
{\cal U}_\nu(t',t_0)[\rho_\alpha^s \otimes \rho^D_{in}(t_a)]{\cal U}_\nu^\dag(t',t_0)=|\tilde{u}_\alpha^s(t')\rangle  \langle \tilde{u}_\alpha^s(t')|\otimes \rho^D_{\alpha}(T).
\label{14}
\end{equation}
The density matrix $\rho^D_{\alpha}(T)$ in Eqs. (\ref{13}) and (\ref{14}) is defined in a space ${\cal G}_D$ as
\begin{equation}
\rho^D_{\alpha}(T)=\int |\Phi_{\alpha}(Z,T)\rangle {\rm Q}(Z,Z')\langle \Phi_{\alpha}(Z',T)| d\mu(Z,Z'),
\label{15}
\end{equation}
with $d\mu(Z,Z')=d\mu(Z)d\mu(Z')$, $\rho^D_{\alpha}(T)\equiv \rho^D_{\alpha}(t',t_0,t_a)$ and ${\rm Tr}\rho^D_\alpha(T)=1$.
Equations (\ref{13}) and (\ref{14}) show that the density matrix $\rho^D_{\alpha}(T)$ describes a system MD+ES where MD indicates the eigenvalue $\lambda_\alpha$ at $t'\geq 0$. 
However, to be rigorous, this assertion should be postulated. 

{\it Postulate 3: The system MD+ES after measuring degenerate or nondegenerate observable ${\cal A}$ is described by density matrix $\rho^D_{\alpha}(T)$ given in Eq. (\ref{15}). This density matrix 
is defined in a space ${\cal G}_D={\cal G}_M\otimes{\cal G}_E$. }
\medskip

Thus, the conditional density matrix of MD is given by ${\rm w}_{\alpha}(T)={\rm Tr}_E[\rho^D_{\alpha}(T)]$ in a space ${\cal G}_M$ as long as the initial point $t_a$ and the final point $t_b$ of measuring
process are fixed. The trace ${\rm Tr}_E$ is computed in a space ${\cal G}_E$ of an environment system. The MD in the state ${\rm w}_{\alpha}(T)$ indicates the eigenvalue $\lambda_\alpha$ at $t'=t-t_b\geq 0$.

The set of measurements of quantum system where MD indicates different eigenvalues $\lambda_\alpha$ is called the full ensemble ${\cal E}$ of measurements of an observable ${\cal A}$.
The subset of measurements where MD indicates only given eigenvalue $\lambda_\alpha$ is called the subensemble ${\cal E}_\alpha$ of measurements of an observable ${\cal A}$.

Note that the conditional density matrix (CDM) describing the measuring process in the space ${\cal G}_Q\otimes{\cal G}_M$ with fixed points $t_a$ and $t_b$ is given by equation
${\rm W}(T)={\rm Tr}_E\{{\cal U}_\nu(t',t_0)[\rho(t_a) \otimes \rho^D_{in}(t_a)]{\cal U}_\nu^\dag(t',t_0)\}$, 
where $\rho(t_a)$ is the density matrix of measurable quantum system, $\rho^D_{in}(t_a)$ is given in Eq. (\ref{7}) and ${\rm W}(T)\equiv {\rm W}(t',t_0,t_a)$. 

{\it Postulate 4: The measurement of an observable ${\cal A}$ with a discrete non-degenerate or degenerate spectrum leads to irreversible evolution of combined subsystem QS+MD.
The conditional density matrix ${\rm W}(T)$ has the form,
\begin{equation}
{\rm W}(T)= \sum_\alpha P_{\alpha}(t_a)\rho_\alpha(t',t_a) \otimes {\rm w}_{\alpha}(T),
\label{16}
\end{equation}
with ${\rm w}_{\alpha}(T)={\rm Tr}_E[\rho^D_{\alpha}(T)]$.}
\medskip

In Eq. (\ref{16}) $\rho_\alpha(t',t_a)$ is the density matrix in a space ${\cal G}_Q$ describing the subensemble ${\cal E}_\alpha$ of QS and $P_{\alpha}(t_a)\equiv P(\lambda_\alpha,t_a)$ is the probability that MD indicates the eigenvalue $\lambda_\alpha$ for the full ensemble ${\cal E}$ of measurements of an observable ${\cal A}$. The probability $P_{\alpha}(t_a)$ and the density matrix $\rho_\alpha(t',t_a)$ satisfy to normalization conditions $\sum_\alpha P_{\alpha}(t_a)=1$ and ${\rm Tr}\rho_\alpha(t',t_a)=1$. The full ensemble ${\cal E}$ of QS at $t'>0$ is described by density matrix $\tilde{\rho}(t',t_a)=\sum_\alpha P_\alpha(t_a) \rho_\alpha(t',t_a)$.

Note that the density matrices $\tilde{\rho}(t',t_a)$ and $\rho_\alpha(t',t_a)$ depend on $t'=t-t_b$, and they do not depend explicitly on $t_b$. 
The evolution of the combined system QS+MD postulated in Eq. (\ref{16}) is irreversible due to the trace operator ${\rm Tr}_E$ in definition of the density matrices ${\rm W}(T)$ and ${\rm w}_{\alpha}(T)$. We emphasize that one can consider the formulated Postulates as a rigorous definition of the MD and measuring process.

If the observable ${\cal A}$ is non-degenerate then Eq. (\ref{11}) in Postulate  2 leads to equation, 
\begin{eqnarray}
{\rm Tr}_E\{{\cal U}_\nu(t',t_0)[\rho(t_a) \otimes \rho^D_{in}(t_a)]{\cal U}_\nu^\dag(t',t_0)\}~~~~~~~~~~~
\nonumber\\ \noalign{\vskip3pt} 
= \sum_{\alpha }  \sum_{\beta } \langle u_\alpha|\rho(t_a)|u_\beta\rangle |u_\alpha(t')\rangle  \langle u_\beta(t')| \otimes {\rm w}_{\alpha\beta}(T),~~~~~
\label{17}
\end{eqnarray}
where ${\rm w}_{\alpha\beta}(T)={\rm Tr}_E[\rho^D_{\alpha\beta}(T)]$ and the operator $\rho^D_{\alpha\beta}(T)$ is given by
\begin{equation}
\rho^D_{\alpha\beta}(T)=\int |\Phi_{\alpha}(Z,T)\rangle {\rm Q}(Z,Z')\langle \Phi_{\beta}(Z',T)| d\mu(Z,Z').
\label{18}
\end{equation}
In the case when the observable ${\cal A}$ is degenerate Eq. (\ref{12}) in Postulate  2 leads to equation,
\begin{eqnarray}
{\rm Tr}_E\{{\cal U}_\nu(t',t_0)[\rho(t_a) \otimes \rho^D_{in}(t_a)]{\cal U}_\nu^\dag(t',t_0)\}~~~~~~~~~~~
\nonumber\\ \noalign{\vskip3pt}  
= \sum_{\alpha s}  \sum_{\beta n} \langle u_\alpha^s|\rho(t_a)|u_\beta^n\rangle |\tilde{u}_\alpha^s(t')\rangle  \langle \tilde{u}_\beta^n(t')| \otimes {\rm w}_{\alpha\beta}(T).~~~~~~~
\label{19}
\end{eqnarray}
Thus, Eqs. (\ref{17}), (\ref{18}), (\ref{19}) and Eq. (\ref{16}) in Postulate 4 lead to the following assertion:

The operator $\rho^D_{\alpha\beta}(T)$ given in Eq. (\ref{18}) and the density matrix $\rho^D_{\alpha}(T)$ defined in  Eq. (\ref{15}) satisfy the relation ${\rm w}_{\alpha\beta}(T)=\delta_{\alpha\beta}{\rm w}_{\alpha}(T)$. In an explicit form this relation is given by
\begin{equation}
{\rm Tr}_E[\rho^D_{\alpha\beta}(T)]=\delta_{\alpha\beta}{\rm Tr}_E[\rho^D_{\alpha}(T)],
\label{20}
\end{equation}
where $\delta_{\alpha\beta}$ is the Kronecker symbol.
This Eq. (\ref{20}) means that ${\rm w}_{\alpha\beta}(T)=0$ when $\alpha\neq \beta$. 

Two observables are called compatible if the measurement one of them does not affect on the prediction of the measurement of the other observable. The Postulates formulated in this section lead also to well-known assertion for non-degenerate observables: 
The observables ${\cal A}$ and ${\cal B}$ are compatible if and only if the operators ${\cal A}$ and ${\cal B}$ commute.

\section{Discrete spectrum measurement}

In this section, we consider the main consequences of the formulated postulates when an observable ${\cal A}$ has a discrete degenerate or non-degenerate spectrum. 
Equation (\ref{16}) in Postulate 4 can be written at $t'\geq 0$ as
\begin{equation}
{\rm W} (T)= \sum_\alpha \sigma_\alpha(t',t_a) \otimes {\rm w}_{\alpha}(T),
\label{21}
\end{equation}
where $\sigma_\alpha(t',t_a)=P_{\alpha}(t_a)\rho_\alpha(t',t_a)$ and ${\rm W}(T)$ is the conditional density matrix of the combined system QS+MD,
\begin{equation}
{\rm W}(T)={\rm Tr}_E\{{\cal U}_\nu(t',t_0)[\rho(t_a) \otimes \rho^D_{in}(t_a)]{\cal U}_\nu^\dag(t',t_0)\}.
\label{22}
\end{equation}
Equation (\ref{21}) and relation ${\rm Tr}\rho_\alpha(t',t_a)=1$ (see Postulate 4) yield the equations for the probability $P_{\alpha}(t_a)$ and density matrix $\rho_\alpha(t',t_a)$ as
\begin{equation}
P_\alpha(t_a)= {\rm Tr}\sigma_\alpha(t',t_a),~~~\rho_\alpha(t',t_a)=\frac{\sigma_\alpha(t',t_a)}{{\rm Tr}\sigma_\alpha(t',t_a)},
\label{23}
\end{equation}
where the quantity $P_\alpha(t_a)$ is positive and independent on the time $t'$. The density matrix $\tilde{\rho}(t',t_a)$ describing the full ensemble of quantum system after measurements is given by $\tilde{\rho}(t',t_a)=\sum_\alpha \sigma_\alpha(t',t_a)$. 

{\bf Degenerate spectrum}.

Eqs. (\ref{19}) and (\ref{20}) yield the conditional density matrix ${\rm W} (T)$ of the combined system QS+MD as
\begin{equation}
{\rm W} (T)= \sum_\alpha \sum_{sn} \langle u_\alpha^s|\rho(t_a)|u_\alpha^n\rangle |\tilde{u}_\alpha^s(t')\rangle  \langle \tilde{u}_\alpha^n(t')| \otimes {\rm w}_{\alpha}(T).
\label{24}
\end{equation}
This equation can also be written in the form,
 \begin{equation}
{\rm W}(T)=\sum_\alpha {\cal M}_\alpha(t')  \rho(t_a){\cal M}_\alpha^{\dag}(t')  \otimes {\rm w}_{\alpha}(T),
\label{25}
\end{equation}
where ${\cal M}_\alpha(t')={\rm U}_{\rm Q}(t'){\cal M}_\alpha$ and the operator ${\cal M}_\alpha$ is
\begin{equation}
{\cal M}_\alpha =\sum_{s=1}^{g_\alpha} |\tilde{u}_\alpha^s\rangle \langle u_\alpha^s|= \sum_{s=1}^{g_\alpha} 
\sum_{n=1}^{g_\alpha} \Lambda_\alpha^{sn}|u_\alpha^n \rangle \langle u_\alpha^s|.
\label{26}
\end{equation}
Equations (\ref{21}) and (\ref{25}) yield the following relation $\sigma_\alpha(t',t_a)={\cal M}_\alpha(t')\rho(t_a){\cal M}_\alpha^{\dag}(t')$. Hence the probability $P_{\alpha}(t_a)$ given by Eq. (\ref{23}) is
\begin{equation}
P_{\alpha}(t_a)={\rm Tr}(\rho(t_a){\cal M}_\alpha^{\dag}{\cal M}_\alpha)=\sum_{s=1}^{g_\alpha}\langle u_\alpha^s|\rho(t_a)|u_\alpha^s\rangle. 
\label{27}
\end{equation}
Thus the probability $P_{\alpha}(t_a)$ is independent on the time $t'$ because ${\rm U}_{\rm Q}^\dag(t'){\rm U}_{\rm Q}(t')={\bf 1}_{\rm Q}$. The density matrices $\rho_\alpha(t',t_a)$ and $\tilde{\rho}(t',t_a)$ describing the subensemble ${\cal E}_\alpha$ and  full ensemble ${\cal E}$ of quantum system after measurements are given by
\begin{equation}
\rho_\alpha(t',t_a)=\frac{1}{P_{\alpha}(t_a)}\sum_{sn} \langle u_\alpha^s|\rho(t_a)|u_\alpha^n\rangle |\tilde{u}_\alpha^s(t')\rangle  \langle \tilde{u}_\alpha^n(t')|,
\label{28}
\end{equation}
\begin{equation}
\tilde{\rho}(t',t_a)=\sum_\alpha \sum_{sn} \langle u_\alpha^s|\rho(t_a)|u_\alpha^n\rangle |\tilde{u}_\alpha^s(t')\rangle  \langle \tilde{u}_\alpha^n(t')|.
\label{29}
\end{equation}
Equation (\ref{26}) and the orthonormal conditions 
$\langle \tilde{u}_\alpha^s|\tilde{u}_\alpha^n\rangle=\delta_{sn}$ (see Eq. (\ref{3})) lead to relation,
\begin{equation}
\sum_\alpha {\cal M}_\alpha^{\dag}{\cal M}_\alpha=\sum_\alpha\sum_{s=1}^{g_\alpha} \sum_{n=1}^{g_\alpha}|u_\alpha^s\rangle \langle \tilde{u}_\alpha^s|\tilde{u}_\alpha^n\rangle \langle u_\alpha^n|={\bf 1}_{\rm Q}. 
\label{30}
\end{equation}
Thus Eqs. (\ref{27}) and (\ref{30}) yield the normalization condition for the probability as $\sum_\alpha P_\alpha(t_a)=1$.
We have the relation ${\cal M}_\alpha^{\dag}{\cal M}_\alpha=\sum_{s=1}^{g_\alpha} |u_\alpha^s\rangle \langle u_\alpha^s|={\cal P}_\alpha$ where the projection operator ${\cal P}_\alpha$ is also defined in Eq. (\ref{2}). Hence Eq. (\ref{30}) can be written as $\sum_\alpha{\cal P}_\alpha={\bf 1}_{\rm Q}$ and the probability given in Eq. (\ref{27}) has well known projection form,
\begin{equation}
P_{\alpha}(t_a)={\rm Tr}(\rho(t_a){\cal P}_\alpha)=\sum_k\sum_{s=1}^{g_\alpha} \pi_k | \langle u_\alpha^s|\psi_k(t_a)\rangle|^2,
\label{31}
\end{equation}
where $\rho(t)$ is defined in Eq. (\ref{6}).

{\bf Nondegenerate spectrum}.

 In the case when the spectrum of an observable ${\cal A}$ is non-degenerate, ${\cal A}|u_\alpha\rangle=\lambda_\alpha|u_\alpha\rangle$, Eqs. (\ref{17}) and (\ref{20}) yield the density matrix ${\rm W} (T)$ of the combined system QS+MD as
 \begin{equation}
{\rm W} (T)= \sum_\alpha  \langle u_\alpha|\rho(t_a)|u_\alpha\rangle |u_\alpha(t')\rangle  \langle u_\alpha(t')| \otimes {\rm w}_{\alpha}(T).
\label{32}
\end{equation}
It follows from Eqs. (\ref{21}) and (\ref{32}) that $\sigma_\alpha(t',t_a)=\langle u_\alpha|\rho(t_a)|u_\alpha\rangle |u_\alpha(t')\rangle  \langle u_\alpha(t')|$. Thus Eq. (\ref{23}) yields the probability $P_{\alpha}(t_a)$ as
\begin{equation}
P_{\alpha}(t_a)={\rm Tr}(\rho(t_a){\cal P}_\alpha)=\sum_k \pi_k | \langle u_\alpha|\psi_k(t_a)\rangle|^2,
\label{33}
\end{equation}
where ${\cal P}_\alpha=|u_\alpha\rangle \langle u_\alpha|$ is the projection operator.
The normalization condition $\sum_\alpha P_\alpha(t_a)=1$ follows from the relation $\sum_\alpha {\cal P}_\alpha={\bf 1}_{\rm Q}$.  For an example,
if the initial state of measuring quantum system is a pure state $\rho(t_a)=|\psi(t_a)\rangle \langle \psi(t_a)|$ then the probability that the MD indicates the eigenvalue $\lambda_\alpha$ is $P_{\alpha}(t_a)= |\langle u_\alpha|\psi(t_a)\rangle|^2$.
Equations (\ref{23}) and (\ref{32}) lead to the density matrix $\rho_\alpha(t',t_a)$ which does not depend on the  initial time $t_a$
[$\rho_\alpha(t',t_a)\equiv\rho_\alpha(t')$]. Thus in this case the density matrices $\rho_\alpha(t')$ and $\tilde{\rho}(t',t_a)$ describing the subensemble ${\cal E}_\alpha$ and  full ensemble ${\cal E}$ of quantum system after measurements are given by
\begin{equation}
\rho_\alpha(t')=|u_\alpha(t')\rangle \langle u_\alpha(t')|={\rm U}_{\rm Q}(t')|u_\alpha\rangle \langle u_\alpha|{\rm U}_{\rm Q}^{\dag}(t'),
\label{34}
\end{equation}\begin{equation}
\tilde{\rho}(t',t_a)=\sum_\alpha \langle u_\alpha|\rho(t_a)|u_\alpha\rangle |u_\alpha(t')\rangle \langle u_\alpha(t')|.
\label{35}
\end{equation}

\section{Nondemolition measurement}

In this section, we consider an observable ${\cal A}$ with a discrete degenerate spectrum. By this we introduce 
an orthonormal basis of ket vectors $|v_\alpha\rangle$ as
\begin{equation}
|v_\alpha\rangle=\sum_{s=1}^{g_\alpha} c_\alpha^s|u_\alpha^s\rangle,
\label{36}
\end{equation}
where $c_\alpha^s$ are the complex amplitudes and $|u_\alpha^s\rangle$ are the eigenvectors of ${\cal A}$. 
The orthonormal relations $\langle v_\alpha|v_\beta\rangle=\delta_{\alpha\beta}$ and $\langle u_\alpha^n|u_\beta^s\rangle=\delta_{\alpha\beta}\delta_{ns}$ lead to equation $\sum_{s=1}^{g_\alpha} |c_\alpha^s|^2=1$ for complex amplitudes $c_\alpha^s$.
The orthonormal basis given by Eq. (\ref{36}) yields a linear space ${\cal H}_v$ of ket vectors.
We also introduce an orthonormal basis of vectors $|\tilde{v}_\alpha\rangle$ which are connected to the states $|\tilde{u}_\alpha^s\rangle$ (see Eq. (\ref{4})) as 
\begin{equation}
|\tilde{v}_\alpha\rangle=\sum_{s=1}^{g_\alpha} c_\alpha^s|\tilde{u}_\alpha^s\rangle=\sum_{s=1}^{g_\alpha}\sum_{n=1}^{g_\alpha} c_\alpha^s \Lambda_\alpha^{sn}|u_\alpha^n\rangle,
\label{37}
\end{equation}
where $\langle \tilde{v}_\alpha|\tilde{v}_\beta\rangle=\delta_{\alpha\beta}$. 
It follows from Eq. (\ref{1}) that the vectors $|v_\alpha\rangle$ and $|\tilde{v}_\alpha\rangle$ are the eigenstates of the degenerate observable ${\cal A}$,
\begin{equation}
{\cal A}|v_\alpha\rangle=\lambda_\alpha|v_\alpha\rangle,~~~{\cal A}|\tilde{v}_\alpha\rangle=\lambda_\alpha|\tilde{v}_\alpha\rangle.
\label{38}
\end{equation}

We assume that QS can be prepared at $t=t_a$ as a linear superposition of ket vectors $|v_\alpha\rangle$. Thus, the pure states of QS 
belong to the space ${\cal H}_v$ at $t=t_a$. In this case,      
the density matrix $\rho_v(t)$ of QS at $t=t_a$ has the form,
\begin{equation}
\rho_v(t_a)=\sum_k \pi_k |\chi_k(t_a)\rangle \langle \chi_k(t_a)|,~~|\chi_k(t_a)\rangle=\sum_\alpha B_{k\alpha}|v_\alpha\rangle,
\label{39}
\end{equation}
where $\sum_k \pi_k=1$ ($\pi_k>0$) and $\langle \chi_k(t_a)|\chi_k(t_a)\rangle=1$.
The density matrix $\rho_v(t_a)$ can be written as
\begin{equation}
\rho_v(t_a)=\sum_{\alpha\beta} |v_\alpha\rangle w_{\alpha\beta}(t_a) \langle v_\beta|,
\label{40}
\end{equation}
where the matrix $w_{\alpha\beta}(t_a)=\langle v_\alpha|\rho_v(t_a)|v_\beta\rangle$ is 
\begin{equation}
w_{\alpha\beta}(t_a)=\sum_k \pi_k B_{k\alpha}B_{k\beta}^{*},~~~\sum_\alpha  |B_{k\alpha}|^2=1.
\label{41}
\end{equation}
If an observable has a discrete degenerate spectrum then the density matrix ${\rm W} (T)$ of the combined system QS+MD is given by Eq. (\ref{25}) where $\rho(t_a)=\rho_v(t_a)$. Thus we have the equation, 
\begin{equation}
{\rm W}_v(T)=\sum_\alpha {\cal M}_\alpha(t')  \rho_v(t_a){\cal M}_\alpha^{\dag}(t')  \otimes {\rm w}_{\alpha}(T),
\label{42}
\end{equation}
where the conditional density matrix ${\rm W}_v(T)$ is
\begin{equation}
{\rm W}_v(T)={\rm Tr}_E\{{\cal U}_\nu(t',t_0)[\rho_v(t_a) \otimes \rho^D_{in}(t_a)]{\cal U}_\nu^\dag(t',t_0)\}.
\label{43}
\end{equation}
Equations (\ref{40}) and (\ref{42}) lead to conditional density matrix ${\rm W}_v(T)$ in the form
\begin{equation}
{\rm W}_v (T)= \sum_\alpha \langle v_\alpha|\rho_v(t_a)|v_\alpha\rangle |\tilde{v}_\alpha(t')\rangle  \langle \tilde{v}_\alpha(t')| \otimes {\rm w}_{\alpha}(T),
\label{44}
\end{equation}
where $|\tilde{v}_\alpha(t')\rangle ={\rm U}_{\rm Q}(t')|\tilde{v}_\alpha\rangle$. Note that the operator $\sigma_\alpha(t',t_a)$ defined in Eq. (\ref{21}) is given by
\begin{equation}
\sigma_\alpha(t',t_a)=w_{\alpha\alpha}(t_a)|\tilde{v}_\alpha(t')\rangle  \langle \tilde{v}_\alpha(t')|. 
\label{45}
\end{equation}
Thus, the probability is $P_\alpha(t_a)= {\rm Tr}\sigma_\alpha(t',t_a)$. This leads to probability $P_\alpha(t_a)$ which does not depend on time $t'$:
\begin{equation}
P_\alpha(t_a)=\langle v_\alpha|\rho_v(t_a)|v_\alpha\rangle=\sum_k \pi_k |B_{k\alpha}|^2.
\label{46}
\end{equation}
Equations (\ref{41}) and (\ref{46}) yield the normalization condition for probability as $\sum_\alpha P_\alpha(t_a)=1$. In this case
the density matrix $\rho_\alpha(t',t_a)$ does not depend on initial time $t_a$ [ $\rho_\alpha(t',t_a)\equiv\rho_\alpha(t')$].  
The density matrices $\rho_\alpha(t')$ and $\tilde{\rho}(t',t_a)$ describing the subensemble ${\cal E}_\alpha$ and  full ensemble ${\cal E}$ of quantum system after measurements are given by

\begin{equation}
\rho_\alpha(t')=|\tilde{v}_\alpha(t')\rangle  \langle \tilde{v}_\alpha(t')|={\rm U}_{\rm Q}(t')|\tilde{v}_\alpha\rangle\langle \tilde{v}_\alpha|{\rm U}_{\rm Q}^{\dag}(t'),
\label{47}
\end{equation}
\begin{equation}
\tilde{\rho}(t',t_a)=\sum_\alpha \langle v_\alpha|\rho_v(t_a)|v_\alpha\rangle |\tilde{v}_\alpha(t')\rangle  \langle \tilde{v}_\alpha(t')|.
\label{48}
\end{equation}
The probability given in Eq. (\ref{46}) can also be written as
\begin{equation}
P_{\alpha}(t_a)=\sum_k\pi_k | \langle v_\alpha|\chi_k(t_a)\rangle|^2,
\label{49}
\end{equation}
which is similar to the non-degenerate case.  
Note that the states $|v_\alpha\rangle$ are also the eigenstates of an observable $\tilde{{\cal A}}$ given by
\begin{equation}
\tilde{{\cal A}}=\sum_\alpha \lambda_\alpha \tilde{{\cal P}}_\alpha,~~~\tilde{{\cal A}}|v_\alpha\rangle=\lambda_\alpha|v_\alpha\rangle,
\label{50}
\end{equation}
where $\tilde{{\cal P}}_\alpha=|v_\alpha\rangle \langle v_\alpha|$ is a projector operator.
Hence, Eqs. (\ref{47})--(\ref{49}) describe the measurement of an observable $\tilde{{\cal A}}$ when $\Lambda_\alpha^{sn}=\delta_{sn}$.
It follows from Postulate 4 and Eqs. (\ref{44}) and (\ref{47}) the next assertion:

If the density matrix of QS at $t=t_a$ is given by Eq. (\ref{39}) and MD indicates the eigenvalue $\lambda_\alpha$ then the measurable QS at $t'\geq0$ has a pure state $|\tilde{v}_\alpha(t')\rangle ={\rm U}_{\rm Q}(t')|\tilde{v}_\alpha\rangle$. 
In the case when $\Lambda_\alpha^{sn}=\delta_{sn}$ this pure state is $|v_\alpha(t')\rangle ={\rm U}_{\rm Q}(t')|v_\alpha\rangle$.

Hence if MD indicates the eigenvalue $\lambda_\alpha$ then the measurable QS at $t=t_b$ (straight after measurement) has a pure state $|\tilde{v}_\alpha\rangle$ and in the case when $\Lambda_\alpha^{sn}=\delta_{sn}$ this pure state is $|v_\alpha\rangle$. 
Thus, the measurement does not destroy the superposition of degenerate states given in Eq. (\ref{36}).

\section{Continuous spectrum measurement}

In this section, we consider the measurement of the observables with continuous spectrum. 
It is well-known that the quantum theory based on Hilbert Space ${\cal H }$ leads to mathematical problems for unbounded operators because such operators are not defined on the whole of ${\cal H }$, but only on dense subdomains of ${\cal H }$ that are not invariant under the action of the observables. The non-invariance makes expectation values, uncertainties and commutation relations not well defined on the whole of ${\cal H }$. This problem can be solved by 
an extension of a Hilbert space to a rigged Hilbert space (see Appendix B). 

The position and momentum operators defined on singular states $|{\bf x}\rangle$ and $|{\bf p}\rangle $ are given by
\begin{equation}
{\bf X}=\int {\bf x}|{\bf x}\rangle \langle {\bf x}|d{\bf x},~~~{\bf P}=\int {\bf p}|{\bf p}\rangle \langle {\bf p}|d{\bf p}.
\label{51}
\end{equation}
The closure relations for these singular states are
\begin{equation}
\int |{\bf x}\rangle \langle {\bf x}|d{\bf x}={\bf 1},~~~\int |{\bf p}\rangle \langle {\bf p}|d{\bf p}={\bf 1},
\label{52}
\end{equation}
where it is assumed the singular normalization $\langle {\bf x}'|{\bf x}\rangle=\delta({\bf x}-{\bf x}')$ and $\langle {\bf p}'|{\bf p}\rangle=\delta({\bf p}-{\bf p}')$, and the scalar product of the states $|{\bf x}\rangle$ and $|{\bf p}\rangle $ is $\langle {\bf x}|{\bf p}\rangle=(2\pi\hbar)^{-3/2}\exp[(i/\hbar){\bf p}\cdot{\bf x}]$. These relations yield the following equations for the position and momentum operators,
\begin{equation}
{\bf X}|{\bf x}\rangle={\bf x}|{\bf x}\rangle,~~~{\bf P}|{\bf p}\rangle ={\bf p}|{\bf p}\rangle .
\label{53}
\end{equation}
The singular states $|{\bf x}\rangle$ and $|{\bf p}\rangle $ do not belong to a Hilbert Space ${\cal H }$. We show below that it is possible to define the states $|{\bf x},\varepsilon\rangle$ and $|{\bf p},\varepsilon\rangle$ which belong to a RHS. Moreover, these states are the eigenvectors of the position and momentum observables with a discrete spectrum.

Let a discrete set of the momentum vectors ${\bf p}$ is given by the following relation,
\begin{equation}
{\bf p}=\varepsilon {\bf n}=\varepsilon(n_1,n_2,n_3),~~~~~~n_k=0,\pm1,\pm 2,...~,
\label{54}
\end{equation}
where ${\bf n}$ is a vector with the discrete components and $\varepsilon$ is an arbitrary positive parameter.
We define a cubic cell $v_{\bf p}$ with a discrete momentum ${\bf p}$ as
\begin{equation}
\left\{{\bf p}'\in v_{\bf p}: \varepsilon\left(n_k-\frac{1}{2}\right)<{\rm p}_k'<\varepsilon\left(n_k+\frac{1}{2}\right)\right\},
\label{55}
\end{equation}
where $k=1,2,3$, and ${\bf p}'$ is an arbitrary momentum vector belonging to a cubic cell $v_{\bf p}$. 
The volume of the cubic cell $v_{\bf p}$ is $\varepsilon^3$. The test function $\tilde{\phi}_\varepsilon({\bf p},{\bf p}')$ of the state $|{\bf p},\varepsilon\rangle $ in the RHS can be chosen as
\begin{equation}
\tilde{\phi}_\varepsilon({\bf p},{\bf p}')=\left\{\begin{array}{ll}
\varepsilon^{-3/2}  & \mbox{if ${\bf p}'\in v_{\bf p}$}\\
0 & \mbox{otherwise}
\end{array}
\right. 
\label{56}
\end{equation}
where ${\bf p}$ is the discrete vector given in Eq. (\ref{54}).
The generalized eigenstate $|{\bf p},\varepsilon\rangle$ in a rigged Hilbert space with a test function $\tilde{\phi}_\varepsilon({\bf p},{\bf p}')$ is 
\begin{equation}
|{\bf p},\varepsilon\rangle =\int \tilde{\phi}_\varepsilon({\bf p},{\bf p}')|{\bf p}'\rangle d{\bf p}'=\varepsilon^{-3/2}\int_{v_{\bf p}} |{\bf p}'\rangle d{\bf p}'. 
\label{57}
\end{equation}
It is shown in Appendix C that these states are orthonormal $\langle{\bf p}',\varepsilon|{\bf p},\varepsilon\rangle=\delta_{ {\bf p}'{\bf p}}$ where $\delta_{ {\bf p}'{\bf p}}$ is the Kronecker symbol. The set of states $|{\bf p},\varepsilon\rangle$ is also complete asymptotically in the RHS at $\varepsilon\rightarrow 0$ (see Appendix C): 
\begin{equation}
\sum_{{\bf p}}|{\bf p},\varepsilon\rangle\langle{\bf p},\varepsilon|=\int |{\bf p}\rangle \langle{\bf p}|d{\bf p} ={\bf 1}.
\label{58}
\end{equation}
Hence, the decomposition of the density matrix $\rho(t)$ in a rigged Hilbert space at $\varepsilon\rightarrow 0$ is given by 
\begin{equation}
\rho(t)=\sum_{{\bf p}'}\sum_{{\bf p}''}  \langle {\bf p}',\varepsilon|\rho(t)|{\bf p}'',\varepsilon\rangle |{\bf p}',\varepsilon\rangle \langle{\bf p}'',\varepsilon|.
\label{59}
\end{equation}

Eq. (\ref{11}) of Postulate 2 for the momentum state $|{\bf p},\varepsilon\rangle$ in the RHS has the form,
\begin{equation}
{\cal U}_\nu(t',t_0)|{\bf p},\varepsilon\rangle  \otimes|\Phi_{in}(Z,t_a)\rangle =|{\bf p},\varepsilon;t'\rangle  \otimes
|\Phi_{\bf p}^\varepsilon(Z,T)\rangle.
\label{60}
\end{equation}
The states $|{\bf p},\varepsilon;t'\rangle$ and $|\Phi_{\bf p}^\varepsilon(Z,T)\rangle$ are
\begin{equation}
|{\bf p},\varepsilon;t'\rangle={\rm U}_{\rm Q}(t')|{\bf p},\varepsilon\rangle,~|\Phi_{\bf p}^\varepsilon(Z,T)\rangle={\rm U}_{\rm D}(t')
|\Phi_{\bf p}^\varepsilon(Z,T_0)\rangle,
\label{61}
\end{equation}
where $|\Phi_{\bf p}^\varepsilon(Z,T_0)\rangle=|\Phi_{\bf p}^\varepsilon(Z,0,t_0,t_a)\rangle$.
The density matrix $\rho^D_{\bf p}(\varepsilon;T)$ of the system MD+ES in the subspace ${\cal G}_D$ (see Eq. (\ref{15})) is given by 
\begin{equation}
\rho^D_{\bf p}(\varepsilon;T)=\int |\Phi_{\bf p}^\varepsilon(Z,T)\rangle {\rm Q}(Z,Z')\langle \Phi_{\bf p}^\varepsilon(Z',T)| d\mu(Z,Z'),
\label{62}
\end{equation}
where ${\rm Tr}\rho^D_{\bf p}(\varepsilon;T)=1$. In this case
Eq. (\ref{20}) has the form,
\begin{equation}
{\rm Tr}_E[\rho^D_{{\bf p}{\bf q}}(\varepsilon;T)]=\delta_{{\bf p}{\bf q}}{\rm Tr}_E[\rho^D_{{\bf p}}(\varepsilon;T)],
\label{63}
\end{equation}
where the operator $\rho^D_{{\bf p}{\bf q}}(\varepsilon;T)$ is given by
\begin{equation}
\rho^D_{{\bf p}{\bf q}}(\varepsilon;T)=\int |\Phi_{\bf p}^\varepsilon(Z,T)\rangle {\rm Q}(Z,Z')\langle \Phi_{\bf q}^\varepsilon(Z',T)| d\mu(Z,Z').
\label{64}
\end{equation}
Eqs. (\ref{59})-(\ref{64}) lead to relation, 
\begin{equation}
{\rm W}_\varepsilon (T) = \sum_{\bf p}  \langle {\bf p},\varepsilon|\rho(t_a)|{\bf p},\varepsilon\rangle |{\bf p},\varepsilon;t'\rangle \langle{\bf p},\varepsilon;t'|\otimes {\rm w}_{\bf p}(\varepsilon;T),
\label{65}
\end{equation}
which is analogous to Eq. (\ref{32}).
Here ${\rm W}_\varepsilon (t)$ is the conditional density matrix of the combined system QS+MD in the space ${\cal G}_Q\otimes{\cal G}_M$  (see Eq. (\ref{22})) and ${\rm w}_{\bf p}(\varepsilon;T)={\rm Tr}_E[\rho^D_{{\bf p}}(\varepsilon;T)]$. In this case Eq. (\ref{16}) of Postulate 4 has the form,
\begin{equation}
{\rm W}_\varepsilon (T) =\sum_{\bf p} P_{\bf p}(\varepsilon;t_a) \rho_{{\bf p}}(\varepsilon;t',t_a)\otimes {\rm w}_{\bf p}(\varepsilon;T),
\label{66}
\end{equation}
where $P_{\bf p}(\varepsilon;t_a)$ is the probability to measure the momentum ${\bf p}$ at $t=t_a$. 
We define the operator $\sigma_{\bf p}(\varepsilon;t',t_a)$ (see Eq. (\ref{21})) as
\begin{equation}
\sigma_{\bf p}(\varepsilon;t',t_a)=P_{\bf p}(\varepsilon;t_a)\rho_{\bf p}(\varepsilon;t',t_a),
\label{67}
\end{equation}
where $P_{\bf p}(\varepsilon;t_a)={\rm Tr}\sigma_{\bf p}(\varepsilon;t',t_a)$. 
Equations (\ref{60})-(\ref{67}) lead to relation,
\begin{equation}
\sigma_{\bf p}(\varepsilon;t',t_a)= \langle {\bf p},\varepsilon|\rho(t_a)|{\bf p},\varepsilon\rangle |{\bf p},\varepsilon;t'\rangle \langle{\bf p},\varepsilon;t'|.
\label{68}
\end{equation}
It follows from this equation that the probability $P_{\bf p}(\varepsilon;t_a)$ is given by
\begin{equation}
P_{\bf p}(\varepsilon;t_a)={\rm Tr}\sigma_{\bf p}(\varepsilon;t',t_a)= \langle {\bf p},\varepsilon|\rho(t_a)|{\bf p},\varepsilon\rangle.
\label{69}
\end{equation}
Equations (\ref{57}) and (\ref{69}) yield the asymptotic relation (at $\varepsilon\rightarrow 0$) as
$P_{\bf p}(\varepsilon;t_a)\sim\varepsilon^3\langle{\bf p}|\rho(t_a)|{\bf p}\rangle$ where $\varepsilon^3$ is the volume of the infinitesimal cubic cell $v_{\bf p}$. Thus, the probability distribution to measure the momentum $\bf p$ of the particle at $t=t_a$ is
\begin{equation}
P({\bf p},t_a)=\langle{\bf p}|\rho(t_a)|{\bf p}\rangle=\sum_k \pi_k|\langle{\bf p}|\psi_k(t_a)\rangle|^2.
\label{70}
\end{equation}
The normalization condition $\sum_{\bf p} P_{\bf p}(\varepsilon;t_a)=1$ yields $\int P({\bf p},t_a)d{\bf p}=1$ in the limit 
$\varepsilon\rightarrow 0$. 
Hence, if the quantum system before measurement has a pure state 
$\rho(t)=|\psi(t)\rangle\langle\psi(t)|$, then the probability distribution is $P({\bf p},t_a)=|\langle{\bf p}|\psi(t_a)\rangle|^2$.
It follows from Eqs. (\ref{67})-(\ref{69}) that the subensemble ${\cal E}_{\bf p}$ of quantum system (after measuring of the momentum ${\bf p}$) is described by density matrix $\rho_{\bf p}(\varepsilon;t')=|{\bf p},\varepsilon;t'\rangle \langle{\bf p},\varepsilon;t'|$ which does not depend on the time $t_a$.
The quantum system after momentum measurements is described by density matrix,
\begin{equation}
\tilde{\rho}(\varepsilon;t',t_a)=\sum_{\bf p}  \langle {\bf p},\varepsilon|\rho(t_a)|{\bf p},\varepsilon\rangle |{\bf p},\varepsilon;t'\rangle \langle{\bf p},\varepsilon;t'|.
\label{71}
\end{equation}
 
The conditional density matrix describing the position measurement of quantum system in the space ${\cal G}_Q\otimes{\cal G}_M$  is completely analogous to Eq. (\ref{65}). It has the form,
\begin{equation}
{\rm W}_\varepsilon (T) =\sum_{\bf x} \langle {\bf x,\varepsilon}|\rho(t_a)|{\bf x,\varepsilon}\rangle|{\bf x},\varepsilon;t'\rangle \langle{\bf x},\varepsilon;t'|
\otimes {\rm w}_{\bf x}(\varepsilon;T),
\label{72}
\end{equation}
where ${\rm w}_{\bf x}(\varepsilon;T)={\rm Tr}_E[\rho^D_{{\bf x}}(\varepsilon;T)]$. 
The derivation of this equation is based on Postulate 2 and definition of the state $|{\bf x},\varepsilon\rangle$ given in Appendix C.
Equation (\ref{72}) and Postulate 4 yield the position probability of the particle in a RHS as
$P_{\bf x}(\varepsilon;t_a)=\langle {\bf x},\varepsilon|\rho(t_a)|{\bf x},\varepsilon\rangle$. Hence, the probability distribution to measure the position $\bf x$ of the particle at $t=t_a$ is
\begin{equation}
P({\bf x},t_a)=\langle{\bf x}|\rho(t_a)|{\bf x}\rangle=\sum_k \pi_k|\langle{\bf x}|\psi_k(t_a)\rangle|^2.
\label{73}
\end{equation}
The derivation of this equation is analogous to Eq. (\ref{70}). 
The normalization condition $\sum_{\bf x} P_{\bf x}(\varepsilon;t_a)=1$ yields $\int P({\bf x},t_a)d{\bf x}=1$ in the limit  
$\varepsilon\rightarrow 0$.  
The subensemble of quantum system (after measuring of the position ${\bf x}$) is described by density matrix $\rho_{\bf x}(\varepsilon;t')=|{\bf x},\varepsilon;t'\rangle \langle{\bf x},\varepsilon;t'|$ which does not depend on the time $t_a$. The full ensemble ${\cal E}$ of quantum system after position measurements is given by density matrix,
\begin{equation}
\tilde{\rho}(\varepsilon;t',t_a)=\sum_{\bf x} \langle {\bf x,\varepsilon}|\rho(t_a)|{\bf x,\varepsilon}\rangle|{\bf x},\varepsilon;t'\rangle \langle{\bf x},\varepsilon;t'|,
\label{74}
\end{equation}
where $|{\bf x},\varepsilon;t'\rangle={\rm U}_{\rm Q}(t')|{\bf x},\varepsilon\rangle$ and the state $|{\bf x},\varepsilon\rangle$ is defined in a RHS (see Appendix C).

\section{Non-ideal measurements}

We consider in this section the quantum-measurement theory with non-ideal
initial conditions. The density matrix ${\rm W}(T)\equiv {\rm W}(t',t_0,t_a)$ of the combined system QS+MD  
is defined in Eq. (\ref{22}) where $\rho(t_a)$ is a density matrix of measurable QS at $t=t_a$. Here $t_a$ is the initial time-point of measuring procedure. However, the starting points when
the  MD is involved to measuring process can be differ from $t_a$. Moreover, these points can not be 
defined because the initial state of measuring device given by density matrix $\rho^D_{in}(t_a)$ is metastable (see Postulate 1). Note that difficulties in definition of the starting point of measuring process are discussed in Ref. \cite{SD}. The uncertainty in definition of such time-points we refer to non-ideal initial conditions for measuring process. In general case, the starting time-points of measuring process for the full ensemble of measurements can be described by
probability distribution. We define the survival probability distribution ${\rm P}(t,t_a)=\tilde{{\rm P}}(t-t_a)$ 
which is a probability distribution that $t$ ($t\geq t_a$) is the starting point of measuring process. Thus, we distinguish the initial time-point $t_a$ and starting points of measuring process. 
It is assumed that the function 
 $\tilde{{\rm P}}(t)$ is zero at $t>\tau_0$ where $\tau_0$ is a characteristic time of measuring process. Hence, the normalization condition has the form $\int_0^{\infty} \tilde{{\rm P}}(t)dt=\int_0^{\tau_0} \tilde{{\rm P}}(t)dt=1$. We define the reduced density matrix $\rho_r(t_a)$ by relation,
\begin{equation}
\rho_r(t_a)=\int_{t_a}^{t_a+\tau_0} \rho(t){\rm P}(t,t_a)dt=\int_0^{\tau_0} \rho(t_a+t')\tilde{{\rm P}}(t')dt',
\label{75}
\end{equation}
where $\rho(t)={\rm U}_{\rm Q}(t-t_a)\rho(t_a){\rm U}_{\rm Q}^{\dag}(t-t_a)$. This definition yields the normalization condition ${\rm Tr}\rho_r(t_a)=1$.  Equation (\ref{75}) can be written in the form,
 \begin{equation}
\rho_r(t_a)={\rm lim}_{N\rightarrow \infty} \sum_{i=1}^Nw^{(i)}\rho^{(i)},
\label{76}
\end{equation}
where $w^{(i)}=\tilde{{\rm P}}(t_i)\Delta_N$,  $\rho^{(i)}=\rho(t_a+t_i)$, $\Delta_N=\tau_0/N$, and $t_i=(i-1/2)\Delta_N$. It follows by definition that ${\rm Tr}\rho^{(i)}=1$ and ${\rm lim}_{N\rightarrow \infty}\sum_{i=1}^Nw^{(i)}=1$.
Thus, the quantity $w^{(i)}$ is the probability and $\rho^{(i)}$ is an appropriate partial density matrix. Let the measurable observable is degenerate then the density matrix describing 
the subensemble (see Eq. (\ref{29})) connected to partial density matrix $\rho^{(i)}$ is
\begin{equation}
\tilde\rho^{(i)}(t',t_a)= \sum_\alpha \sum_{sn} \langle u_\alpha^s|\rho^{(i)}|u_\alpha^n\rangle |\tilde{u}_\alpha^s(t')\rangle  \langle \tilde{u}_\alpha^n(t')|.
\label{77}
\end{equation}
Hence, the density matrix describing the full ensemble of QS for non-ideal initial conditions is
$\tilde{\rho}(t',t_a)={\rm lim}_{N\rightarrow \infty} \sum_{i=1}^Nw^{(i)}\tilde\rho^{(i)}(t')$. This relation and Eqs. (\ref{76}) and (\ref{77}) lead to the density matrix describing the full ensemble of QS for non-ideal initial conditions as
\begin{equation}
\tilde{\rho}(t',t_a)= \sum_\alpha \sum_{sn} \langle u_\alpha^s|\rho_r(t_a)|u_\alpha^n\rangle |\tilde{u}_\alpha^s(t')\rangle  \langle \tilde{u}_\alpha^n(t')|.
\label{78}
\end{equation}
Hence, in this case, the probability $P_{\alpha}(t_a)$ is given by
\begin{equation}
P_{\alpha}(t_a)={\rm Tr}(\rho_r(t_a){\cal P}_\alpha)=\sum_{s=1}^{g_\alpha} \langle u_\alpha^s|\rho_r(t_a)|u_\alpha^s\rangle.
\label{79}
\end{equation}
Eqs. (\ref{78}) and (\ref{79}) demonstrate that the survival effect can be taken into account by  the change of the density matrix $\rho(t_a)$ of measurable QS to a reduced density matrix $\rho_r(t_a)$. This statement we formulate by Postulate 5 which one may consider as a definition of the non-ideal measurements.

{\it Postulate 5: The measurements with non-ideal
initial conditions are described by Postulates 1--4 with the change of the density 
matrix $\rho(t_a)$ of measurable quantum system to a reduced density matrix $\rho_r(t_a)$.}   
\medskip

The reduced density matrix is given in Eq. (\ref{75}) where $\tilde{{\rm P}}(t)$ is a survival probability distribution.
Hence, the conditional density matrix ${\rm W}(T)$ of the combined system QS+MD has the form ${\rm W}(T)={\rm Tr}_E\{{\cal U}_\nu(t',t_0)[\rho_r(t_a) \otimes \rho^D_{in}(t_a)]{\cal U}_\nu^\dag(t',t_0)\}$.
Thus, the probability to measure the eigenvalue $\lambda_\alpha$ for degenerate
and non-degenerate observable ${\cal A}$ are given by equations $P_{\alpha}(t_a)=\sum_{s=1}^{g_\alpha} \langle u_\alpha^s|\rho_r(t_a)|u_\alpha^s\rangle $ and $P_{\alpha}(t_a)=\langle u_\alpha|\rho_r(t_a)|u_\alpha\rangle$ respectively, and
the time evolution of density matrix $\rho(t)$ in Eq. (\ref{75}) is defined by the Hamiltonian ${\rm H}_{\rm Q}$. The eigenvalues and eigenvectors of QS follow from equation:  
\begin{equation}
{\rm H}_{\rm Q} |\phi_n\rangle=\epsilon_n|\phi_n\rangle,
\label{80}
\end{equation}
where we assume that the eigenstates are orthonormal, $\langle \phi_n|\phi_m\rangle=\delta_{nm}$.
Hence, the density matrix $\rho(t)$ can be written as 
\begin{equation}
\rho(t)=\sum_{nm} \rho_{nm}(t_a) |\phi_n\rangle \langle \phi_m|\exp{[-i(\omega_n-\omega_m)(t-t_a)]},
\label{81}
\end{equation}
where $\omega_n=\epsilon_n/\hbar$ and $\rho_{nm}(t_a)=\langle \phi_n|\rho(t_a)|\phi_m\rangle$.

Eqs. (\ref{75}) and (\ref{81}) lead to reduced density matrix,
\begin{equation}
\rho_r(t_a)=\sum_{nm}\rho_{nm}(t_a)q_{nm}|\phi_n\rangle \langle \phi_m|.
\label{82}
\end{equation}
The matrices $\rho_{nm}(t_a)$ and $q_{nm}$ are given by
\begin{equation}
\rho_{nm}(t_a)=\sum_k \pi_k \langle \phi_n|\psi_k(t_a)\rangle \langle \psi_k(t_a)|\phi_m\rangle,
\label{83}
\end{equation}
\begin{equation}
q_{nm}=\int_0^{\tau_0} \exp{(-i\omega_{nm}t)}\tilde{{\rm P}}(t)dt,
\label{84}
\end{equation}
with $\omega_{nm}=\omega_n-\omega_m$. Note that
one can put $\tau_0=\infty $ in Eq. (\ref{84}) because it is assumed that the function 
 $\tilde{{\rm P}}(t)$ is zero at $t>\tau_0$. The simplest form of the survival probability distribution is given by
\begin{equation}
\tilde{{\rm P}}(t)=\gamma\exp{(-\gamma t)},
\label{85}
\end{equation}
where $\tau=1/\gamma$ and $\tau\ll \tau_0$.
This survival distribution is similar to photoelectron waiting-time distribution for coherent light \cite{VS,CSR}.
In this case, we have the relation $q_{nm}=(1+i\omega_{nm}\tau)^{-1}$.
More general survival distribution is given by
the Gamma distribution, 
\begin{equation}
\tilde{{\rm P}}(t)=\frac{\gamma^s}{\Gamma(s)} t^{s-1}\exp{(-\gamma t)},
\label{86}
\end{equation}
where $\gamma=1/\tau$ ($\tau\ll \tau_0$), $\Gamma(s)$ is the Gamma function and $s\geq 1$. In this case, the matrix $q_{nm}$ is given by
\begin{equation}
q_{nm}=\left(\frac{1}{1+i\omega_{nm}\tau}\right)^s.
\label{87}
\end{equation}
Note that the survival distribution in Eq. (\ref{85}) is a particular case of the Gamma distribution
with $s=1$.

The probability $P_\alpha(t_a)$ for non-ideal measurements of degenerate observable follows from Eq. (\ref{27}) and Postulate 5. In this case, we have the following relation $P_{\alpha}(t_a)={\rm Tr}(\rho_r(t_a){\cal P}_\alpha)$ which yields
\begin{equation}
P_{\alpha}(t_a)=\sum_{s=1}^{g_\alpha} \sum_{nm} \rho_{nm}(t_a)q_{nm}\langle u_\alpha^s|\phi_n\rangle \langle \phi_m|u_\alpha^s\rangle.
\label{88}
\end{equation}

We consider below the case when the Hamiltonian of the measurable QS is ${\rm H}_{\rm Q}={\bf P}^2/2m$. In this case, Eq. (\ref{80}) has the form,
\begin{equation}
{\rm H}_{\rm Q} |{\bf p},\varepsilon\rangle=E({\bf p}) |{\bf p},\varepsilon\rangle,~~~E({\bf p})=\frac{{\bf p}^2}{2m},
\label{89}
\end{equation}
where the state $|{\bf p},\varepsilon\rangle$ is given by Eq. (\ref{57}).
The reduced density matrix $\rho_r(t_a)$ defined in the RHS at $\varepsilon\rightarrow 0$ is
\begin{equation}
\rho_r(t_a)=\sum_{{\bf p}'}\sum_{{\bf p}''}  \langle {\bf p}',\varepsilon|\rho(t_a)|{\bf p}'',\varepsilon\rangle q({\bf p}',{\bf p}'')|{\bf p}',\varepsilon\rangle \langle{\bf p}'',\varepsilon|,
\label{90}
\end{equation}
The function $q({\bf p}',{\bf p}'')$ is given by Eq. (\ref{84}) as
\begin{equation}
q({\bf p}',{\bf p}'')=\int_0^{\tau_0}\exp[-i\omega({\bf p}',{\bf p}'')t]\tilde{{\rm P}}(t)dt,
\label{91}
\end{equation}
with $\omega({\bf p}',{\bf p}'')=\omega({\bf p}')-\omega({\bf p}'')$ and $\omega({\bf p})=E({\bf p})/\hbar$. 
In this case, Eq. (\ref{65}) with the change $\rho(t_a)\rightarrow \rho_r(t_a)$ yields 
\begin{equation}
{\rm W}_\varepsilon (T) = \sum_{\bf p}  \langle {\bf p},\varepsilon|\rho_r(t_a)|{\bf p},\varepsilon\rangle |{\bf p},\varepsilon;t'\rangle \langle{\bf p},\varepsilon;t'|\otimes {\rm w}_{\bf p}(\varepsilon;T),
\label{92}
\end{equation}
where the reduced density matrix $\rho_r(t_a)$ is given by Eq. (\ref{90}). 
It follows from Eq. (\ref{90}) the following relations $\langle {\bf p},\varepsilon|\rho_r(t_a)|{\bf p},\varepsilon\rangle=\langle {\bf p},\varepsilon|\rho(t_a)|{\bf p},\varepsilon\rangle$ and $\langle {\bf p}|\rho_r(t_a)|{\bf p}\rangle=\langle {\bf p}|\rho(t_a)|{\bf p}\rangle$.
The probability distribution $P({\bf p},t_a)$ follows from Eq. (\ref{70}) and Postulate 5. This procedure and relation $\langle {\bf p}|\rho_r(t_a)|{\bf p}\rangle=\langle {\bf p}|\rho(t_a)|{\bf p}\rangle$ lead to equation for the momentum probability distribution as
\begin{equation}
P({\bf p},t_a)=\langle{\bf p}|\rho_r(t_a)|{\bf p}\rangle=\langle{\bf p}|\rho(t_a)|{\bf p}\rangle.
\label{93}
\end{equation}
The subensemble of quantum system (after measurement of the momentum ${\bf p}$) is described by the density matrix $\rho_{\bf p}(\varepsilon;t')=|{\bf p},\varepsilon;t'\rangle \langle{\bf p},\varepsilon;t'|$ and the full ensemble is described by the density matrix,
\begin{equation}
\tilde{\rho}(\varepsilon;t',t_a)=\sum_{\bf p}  \langle {\bf p},\varepsilon|\rho(t_a)|{\bf p},\varepsilon\rangle |{\bf p},\varepsilon;t'\rangle \langle{\bf p},\varepsilon;t'|.
\label{94}
\end{equation}
Hence, if the Hamiltonian of QS is given by ${\rm H}_{\rm Q}={\bf P}^2/2m$ then the ideal and non-ideal initial conditions (Postulate 5) yield the same momentum probability distribution $P({\bf p},t_a)$ given in Eq. (\ref{93}).
This is also correct in a more general case when the eigenvectors $|{\bf p},\alpha,\varepsilon\rangle$ of free Hamiltonian have an additional set $\alpha$ of quantum numbers. Moreover, this assertion is valid for relativistic particles. Hence, the ideal and non-ideal initial conditions lead to the same differential cross sections of particles.

\section{Survival effect}

In this section, we consider the position measurement of quantum system with non-ideal initial conditions. 
Let the Hamiltonian of QS is given by ${\rm H}_{\rm Q}={\bf P}^2/2m$. In this case, the position measurements of quantum system with non-ideal initial conditions are described by Eq. (\ref{72}) and Postulate 5 as
\begin{equation}
	{\rm W}_\varepsilon (T) =\sum_{\bf x} \langle {\bf x,\varepsilon}|\rho_r(t_a)|{\bf x,\varepsilon}\rangle|{\bf x},\varepsilon;t'\rangle \langle{\bf x},\varepsilon;t'|
	\otimes {\rm w}_{\bf x}(\varepsilon;T).
	\label{95}
\end{equation}
Equation (\ref{73}) and Postulate 5 yield the probability distribution as $P({\bf x},t_a)=\langle{\bf x}|\rho_r(t_a)|{\bf x}\rangle$. This relation and Eq. (\ref{90}) lead to equation,
\begin{eqnarray}
	P({\bf x},t_a)=\sum_{{\bf p}'{\bf p}''} \langle {\bf p}',\varepsilon|\rho(t_a)|{\bf p}'',\varepsilon\rangle 
	\nonumber\\ \noalign{\vskip3pt} \times  q({\bf p}',{\bf p}'')\langle{\bf x}|{\bf p}',\varepsilon\rangle \langle{\bf p}'',\varepsilon|{\bf x}\rangle,
	\label{96}
\end{eqnarray}
where it is assumed that $\varepsilon\rightarrow 0$. It follows from Eq. (\ref{95}) that
the subensemble and the full ensemble of quantum system (after measurement of the position of particle) are described by density matrix $\rho_{\bf x}(\varepsilon;t')=|{\bf x},\varepsilon;t'\rangle \langle{\bf x},\varepsilon;t'|$  and $\tilde{\rho}(\varepsilon;t',t_a)$ as
\begin{equation}
	\tilde{\rho}(\varepsilon,t',t_a)=\sum_{\bf x} \langle {\bf x,\varepsilon}|\rho_r(t_a)|{\bf x,\varepsilon}\rangle|{\bf x},\varepsilon;t'\rangle \langle{\bf x},\varepsilon;t'|.
	\label{97}
\end{equation}
Note that Eq. (\ref{96}) at $\varepsilon \rightarrow 0$ has the form,
\begin{eqnarray}
	P({\bf x},t_a)=\int\int \langle {\bf p}'|\rho(t_a)|{\bf p}''\rangle 
	\nonumber\\ \noalign{\vskip3pt} \times  q({\bf p}',{\bf p}'')\langle{\bf x}|{\bf p}'\rangle \langle{\bf p}''|{\bf x}\rangle d{\bf p}'
	d{\bf p}''.
	\label{98}
\end{eqnarray}
We assume below that a realistic survival distribution can be approximated by the function in Eq. (\ref{86}) with appropriate parameters $\gamma$ and $s$.
In this case, the function $q({\bf p}',{\bf p}'')$ is given by Eq. (\ref{87}) as
\begin{equation}
	q({\bf p}',{\bf p}'')=\left(\frac{1}{1+i\omega({\bf p}',{\bf p}'')\tau}\right)^s,
	\label{99}
\end{equation}
with $\omega({\bf p}',{\bf p}'')=\omega({\bf p}')-\omega({\bf p}'')$ and $\omega({\bf p})=E({\bf p})/\hbar$. 
Equation (\ref{98}) can also be written as
\begin{eqnarray}
	P({\bf x},t_a)=\langle{\bf x}|\rho_r(t_a)|{\bf x}\rangle=\frac{1}{(2\pi\hbar)^3}\int\int \langle {\bf p}'|\rho(t_a)|{\bf p}''\rangle 
	\nonumber\\ \noalign{\vskip3pt} \times q({\bf p}',{\bf p}'') \exp\left[\frac{i}{\hbar}({\bf p}'-{\bf p}'')\cdot {\bf x}\right]d{\bf p}'d{\bf p}''.~~~~~~~~
	\label{100}
\end{eqnarray}
The probability distribution given in Eq. (\ref{100}) differs from the probability distribution defined for the ideal initial conditions because $q({\bf p}',{\bf p}'')\neq 1$.
We call such difference as a survival effect. 

We consider below the survival effect in the first order to small parameter $\epsilon\equiv\tau/\tau_0\ll 1$. The parameter 
$\epsilon$ is small because it is assumed that the function 
$\tilde{{\rm P}}(t)$ is zero at $t>\tau_0$ where $\tau_0$ is a characteristic time of measuring process. In this case, Eq. (\ref{100}) can be essentially simplified. 
Note that Eqs. (\ref{90}) and (99) lead to equation, 
\begin{equation}
	\langle {\bf p}'|\rho_r(t_a)|{\bf p}''\rangle=\frac{\langle {\bf p}'|\rho(t_a)|{\bf p}''\rangle}{[1+i\epsilon\omega({\bf p}',{\bf p}'')\tau_0]^{s}},
	\label{101}
\end{equation}
where $\epsilon\equiv\tau/\tau_0$. The first order to small parameter $\epsilon$ in
Eq. (\ref{101}) yields relation,
\begin{eqnarray}
	\langle {\bf p}'|\rho_r(t_a)|{\bf p}''\rangle=\langle {\bf p}'|\rho(t_a)|{\bf p}''\rangle
	\nonumber\\ \noalign{\vskip3pt}  -is\tau[\omega({\bf p}')-\omega({\bf p}'')]\langle {\bf p}'|\rho(t_a)|{\bf p}''\rangle.
	\label{102}
\end{eqnarray}
This equation can also be written in the form,
\begin{equation}
	\rho_r(t_a)=\rho(t_a)-\frac{is\tau}{\hbar}[{\rm H}_{\rm Q},\rho(t_a)].
	\label{103}
\end{equation}
Equation (\ref{103}) and relation $P({\bf x},t_a)=\langle {\bf x}|\rho_r(t_a)|{\bf x}\rangle$ lead to probability distribution as
\begin{equation}
	P({\bf x,t_a})=\langle {\bf x}|\rho|{\bf x}\rangle+\frac{is\tau}{\hbar}(\langle {\bf x}|\rho{\rm H}_{\rm Q}|{\bf x}\rangle-\langle {\bf x}|{\rm H}_{\rm Q}\rho|{\bf x}\rangle),
	\label{104}
\end{equation}
where $\rho=\rho(t_a)$.
This equation can also be written as 
\begin{equation}
	P({\bf x},t_a)=\langle {\bf x}|\rho(t_a)|{\bf x}\rangle-\frac{2s\tau}{\hbar}{\rm Im}f({\bf x},t_a),
	\label{105}
\end{equation}
where $f({\bf x},t_a)=\langle {\bf x}|\rho(t_a){\rm H}_{\rm Q}|{\bf x}\rangle$. 
The function $f({\bf x},t)$ can be transformed to the form,
\begin{equation}
	f({\bf x},t)=(2\pi\hbar)^{-3/2}\int \langle {\bf x}|\rho(t)|{\bf p}\rangle\frac{{\bf p}^2}{2m}\exp\left(-\frac{i}{\hbar}{\bf p}\cdot {\bf x}\right)d{\bf p}.
	\label{106}
\end{equation}
In the case when the density matrix of QS is given by $\rho(t)=|\psi(t)\rangle \langle \psi(t)|$ the function $f({\bf x},t)$ has the form,
\begin{equation}
	f({\bf x},t)=-\frac{\hbar^2}{2m}\langle{\bf x}|\psi(t)\rangle\nabla^2\int  \langle \psi(t)|{\bf p}\rangle\frac{e^{-(i/\hbar){\bf p}\cdot{\bf x}}}{(2\pi\hbar)^{3/2}}d{\bf p}.
	\label{107}
\end{equation}
Hence, the function $f({\bf x},t)$ can be written as
\begin{equation}
	f({\bf x},t)=-\frac{\hbar^2}{2m}\psi({\bf x},t)\nabla^2\psi^{*}({\bf x},t),
	\label{108}
\end{equation}
where $\psi({\bf x},t)=\langle{\bf x}|\psi(t)\rangle$.
Thus, Eq. (\ref{105}) can be written
\begin{equation}
	P({\bf x},t_a)=|\psi({\bf x},t_a)|^2+\frac{\hbar s\tau}{m}{\rm Im}[\psi({\bf x},t_a)\nabla^2\psi^{*}({\bf x},t_a)].
	\label{109}
\end{equation}
It follows from Eq. (\ref{109}) that the probability distribution has the form $P({\bf x},t_a)=|\psi({\bf x},t_a)|^2$ when the wave function is given by $\psi({\bf x},t_a)=e^{i\alpha}\tilde{\psi}({\bf x},t_a)$. Here $\alpha$ is a real parameter and $\tilde{\psi}({\bf x},t_a)$ is an arbitrary real function.
In the general case Eqs. (\ref{6}) and (\ref{105}) yield the probability distribution as
\begin{eqnarray}
	P({\bf x},t_a)=\sum_k \pi_k|\psi_k({\bf x},t_a)|^2~~~~~~~~
	\nonumber\\ \noalign{\vskip3pt} +\frac{\hbar s\tau}{m}\sum_k \pi_k{\rm Im}[\psi_k({\bf x},t_a)\nabla^2\psi_k^{*}({\bf x},t_a)],
	\label{110}
\end{eqnarray}
which is a generalization of Eq. (\ref{109}). 
Note that Eq. (\ref{103}) leads to normalization condition: ${\rm Tr}\rho_r(t_a)=1$. Hence, the probability distribution in Eqs. (\ref{109}) and (\ref{110}) is  nomalized to unity: $\int P({\bf x},t_a) d{\bf x}=1$. We emphasise that Eqs. (\ref{109}) and (\ref{110}) also follow from Eq. (\ref{100}) in the first order to small parameter $\epsilon$.

Note that in this approximation the function $P({\bf x},t_a)$ in  Eqs. (\ref{109}) and (\ref{110}) can be negative in the region where it is close to zero because these equations are derived in the first order to small parameter
$\epsilon$. However, the absolute value of the function $P({\bf x},t_a)$ is small in such region and one can avoid this non-physical behaviour using a new positive distribution $\tilde{P}({\bf x},t_a)$. Let $G$ is the region where 
the function $P({\bf x},t_a)$ defined in Eq. (\ref{109}) or (\ref{110}) is positive and $Q=\int_G P({\bf x},t_a)d{\bf x}$ is some normalization constant. We define new probability distribution as $\tilde{P}({\bf x},t_a)=Q^{-1}P({\bf x},t_a)$ if ${\bf x}\in G$ and $\tilde{P}({\bf x},t_a)=0$ otherwise. It follows from this definition that $\tilde{P}({\bf x},t_a)\geq 0$ and the normalization condition $\int \tilde{P}({\bf x},t_a)d{\bf x}=1$ is also satisfied. If the condition $Q-1\ll 1$ is satisfied, then  Eqs. (\ref{109}) and (\ref{110}) are correct in the region $G$. In this case,
one can use positive probability distribution $\tilde{P}({\bf x},t_a)$ instead the function $P({\bf x},t_a)$.

As an example, we consider the minimum uncertainty wave packet defined at $t=t_a$ by 
\begin{equation}
	\psi(x,t_a)=\pi^{-1/4}a^{-1/2}\exp\left(\frac{i}{\hbar}p_0x-\frac{x^2}{2a^2}\right).
	\label{111}
\end{equation}
The quantum-measurement theory with ideal
initial conditions (Postulates 1-4) yields for this
wave function the probability distribution $P(x,t_a)=\pi^{-1/2}a^{-1}\exp(-x^2/a^2)$.
Hence, the average values of the position and the momentum of particle are $\langle x\rangle=0$ and $\langle p\rangle=p_0$. 
In this case, the uncertainty relation is given by $\Delta x\Delta p =\hbar/2$ where $\Delta x=\sqrt{\langle x^2\rangle-\langle x\rangle^2}$ and $\Delta p=\sqrt{\langle p^2\rangle-\langle p\rangle^2}$. Note that Eq. (\ref{111}) gives the momentum probability distribution as  
\begin{equation}
	P(p,t_a)=\pi^{-1/2}b^{-1}\exp\left(-\frac{(p-p_0)^2}{b^2}\right),
	\label{112}
\end{equation}
with $b=\hbar/a$. We consider below the non-ideal
initial conditions (see Postulates 5) which lead to survival measurement effect for the wave packet given in Eq. (\ref{111}). 
It follows from Eq. (\ref{93}) that, in this case, the momentum probability distribution $P(p,t_a)$ is given by Eq. (\ref{112}). 
Equation (\ref{109}) yields the position probability distribution as
\begin{equation}
	P(x,t_a)=\pi^{-1/2}a^{-1}\left(1+\frac{2l}{a^2}x\right)\exp\left(-\frac{x^2}{a^2}\right),
	\label{113}
\end{equation}
where $l=s\tau p_0/m$. This equation leads
to the average position of the particle as $\langle x\rangle=l$. The function $P(x,t_a)$ is negative in the interval $x<-a^2/2l$. 
However, in this interval, the dimensionless function $W(\xi)=aP(x,t_a)=\pi^{-1/2}(1+\sqrt{2\epsilon_0}\xi)\exp(-\xi^2)$ with $\xi=x/a$ 
and $\epsilon_0\equiv 2l^2/a^2$ is very close to zero when $\epsilon_0\ll1$ (see Fig. 1). It means that one can put $P(x,t_a)=0$
in the interval $x<-a^2/2l$ when $\epsilon_0\ll1$. 

More accurate procedure is based on the change of the function $P(x,t_a)$ to
the positive probability distribution $\tilde{P}(x,t_a)$ when $\epsilon_0$ is a small parameter. 
First, we define the normalization constant $Q=\int_{x_0}^{+\infty}P(x,t_a)dx$ where $x_0=-a^2/2l$. The condition $Q-1\ll 1$ is satisfied when $\epsilon_0\ll1$ (see Appendix D). In this case, one can define the positive probability distribution $\tilde{P}(x,t_a)$ instead the function $P(x,t_a)$ as
\begin{equation}
\tilde{P}(x,t_a)=\left\{\begin{array}{ll}
Q^{-1}P(x,t_a)  & \mbox{if $x\geq x_0$}\\
0 & \mbox{otherwise}.
\end{array}
\right. 
\label{114}
\end{equation}

\begin{figure}
\includegraphics[width=9cm,trim=0mm 16mm 0mm 0mm]{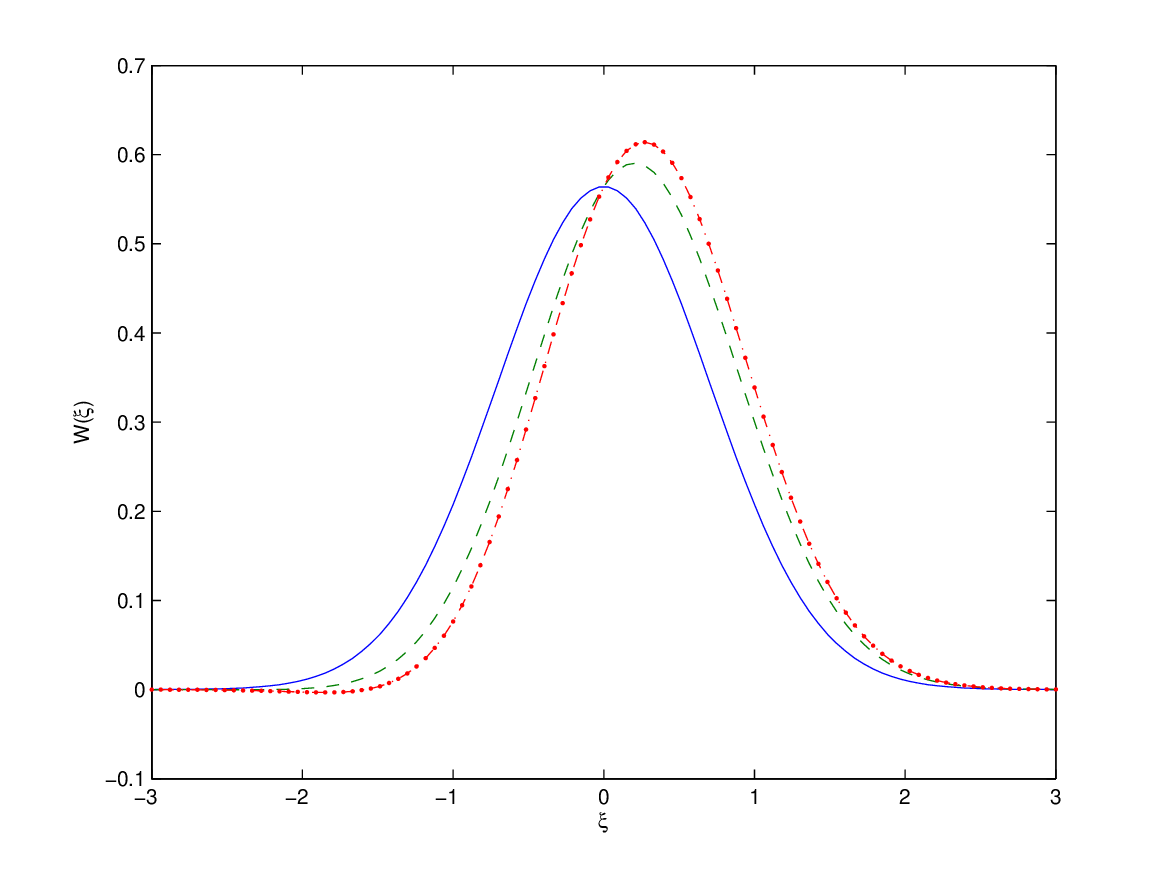}
\caption{Dimensionless distribution $W(\xi)$ for $\epsilon_0=0$ (solid line); $\epsilon_0=0.1$ (dashed line); and $\epsilon_0=0.2$ (dashed-dot line). } 
\label{FIG.}
\end{figure}

Equations (\ref{113}) and (\ref{114}) lead to the same position uncertainty: $\Delta x=\sqrt{a^2/2-l^2}$ when $\epsilon_0\ll1$ (more details are given in Appendix D). The momentum uncertainty follows from Eq. (\ref{112}) as
$\Delta p_{x}=\Delta p=\hbar/\sqrt{2}a$. 
Hence, the survival measurement effect leads to uncertainty relation of the position and momentum of the particle as
\begin{equation}
\Delta x\Delta p_{x}=\frac{\hbar}{2}\left(1-\frac{2l^2}{a^2}\right)^{1/2},~~~l=\frac{s\tau p_0}{m},
\label{115}
\end{equation}
where $2l^2/a^2\ll1$.
Thus, we have the following relation $\Delta x\Delta p_{x}<\hbar/2$ for the minimum uncertainty wave packet. This relation is not consistent with the Heisenberg uncertainty relation $\Delta x\Delta p_{x}\geq \hbar/2$ because it is derived for the non-ideal
initial conditions (see Postulate 5). Hence, the survival effect violates the Heisenberg uncertainty relation. Note  that the survival distributions in Eqs. (\ref{85}) and (\ref{86}) 
become delta-sequences when $\gamma\rightarrow \infty$ (or $\tau\rightarrow 0$). It follows from Eq. (\ref{115}) that in this limiting case the minimum uncertainty wave packet yields the relation $\Delta x\Delta p_{x}=\hbar/2$.  
The above results are found for the case when survival distribution is given by the Gamma distribution. However, it follows from Eqs. (\ref{99}) and (\ref{100}) that the survival effect violates the Heisenberg uncertainty relation for an arbitrary survival distribution if it is not a delta-sequence. Thus, in the general case the Postulate 5 yields the uncertainty relation as 
\begin{equation}
\Delta x\Delta p_{x}\geq\frac{\hbar}{2}A,
\label{116}
\end{equation}  
where the parameter $A$ depends on measuring device or the form of survival probability distribution.
In the example which we consider above the parameter $A$ has the form: $A=\sqrt{1-2(l/a)^2}$ with $2(l/a)^2\ll 1$ (see Eq. (\ref{115})). 
Thus in this case we have $A\neq 1$, however the difference $1-A$ is small. Note that the ideal initial conditions yield relations $\langle p\rangle=p_0$ and $\langle x\rangle=0$ for the minimum uncertainty wave packet. However, the non-ideal initial conditions and the survival distribution in Eq. (\ref{86}) lead to relations $\langle p\rangle=p_0$ and $\langle x\rangle=l$, and hence $\langle x\rangle>0$. These relations can be used for the experimental verification of the survival measurement effect. More complete test of the survival measurement effect is connected with the position probability distribution in Eq. (\ref{114}) and its comparison with the experimental position probability distribution. Note that the experimental position probability distribution for the minimum uncertainty wave packet can be constructed, in principle, by recording the relative measurements. 

\section{Discussion}

In this section we consider in more details different aspects of the survival effect leading to modified (weak) form for the Heisenberg uncertainty relation and the modification of other uncertainty relations. 
  
We note that the survival effect presented in previous section leads to modified (weak) form for the Heisenberg uncertainty relation as
\begin{equation}
\Delta x\Delta p_{x}\gtrsim \frac{\hbar}{2}.
\label{117}
\end{equation} 
This uncertainty relation follows from Eq. (\ref{116}) because we have the following relation $A\simeq 1$.
Using the above treatment of survival effect one also can derive the following  modified Heisenberg uncertainty relation, 
\begin{equation}
\Delta \phi\Delta J_{z}\gtrsim \frac{\hbar}{2},
\label{118}
\end{equation}
where $J_{z}$ is z-component of the orbital angular momentum and $\phi$ is an angle in the $xy$ plane.

We consider here a simple derivation of modified uncertainty relation for energy and time.
However, such derivation of modified uncertainty relation for energy and time needs appropriate definitions. 
Let $\Delta t=\langle\tau\rangle$ is the average time for measuring the change of energy $\delta E=|E-E'|$ where $E$ and $E'$ are the energy of particle before and
after measuring of the particle energy which has the form $\mathcal{E}=p^{2}/2m$. We define $\Delta E=\langle \delta E\rangle$ which can also be written as $\Delta E\simeq \langle \partial \mathcal{E}/\partial p_{x}\rangle \Delta p_{x}=\langle v_{x}\rangle\Delta p_{x}$ where $v_{x}=p_{x}/m$. Thus, using the following relation $\Delta x\simeq \langle v_{x}\rangle \Delta t$ we have the approximate equation:
\begin{equation}
\Delta E\simeq \frac{\Delta x\Delta p_{x}}{\Delta t}.	
\label{119}
\end{equation}  
Combining this important equation with the modified Heisenberg uncertainty relation given in Eq. (\ref{117}) we obtain the following modified uncertainty relation, 
\begin{equation}
\Delta t\Delta E\gtrsim \frac{\hbar}{2}.
\label{120}
\end{equation} 
 
This equation differs from the appropriate
standard form of uncertainty relation for time and energy given as $\Delta t\Delta E\geq \frac{\hbar}{2}$.
Thus, the standard Heisenberg uncertainty relations need the modification presented in Eqs. (\ref{117}), (\ref{118}) and (\ref{120}). This change is important from the theoretical and experimental point of view because it considers the specific influence of measuring device to uncertainty relations. 
We note that W. Heisenberg have derived the uncertainty relations using the Schwarz inequality and appropriate expectation values for a Hermitian operators calculated with the wave function of measuring particles \cite{Hg}. However, in this case the non-ideal initial condition for combined system QS+MD leads to the survival effect.

In the conclusion of this section we discuss some popular terminology connected with the quantum theory of measurement processes. We first turn to the words of P. A. M. Dirac in 1958: ``A measurement always causes the system to jump into an eigenstate of the dynamical variable that is being measured". This means that if the measurement is performed, the final state of measurable system occurs in
some eigenstate $|a\rangle$ of observable  $\cal A$. Thus,
a measurement usually changes the state of measured quantum system. The only exception is  when the initial state is in one of the eigenstate of the observable being measured. The process of such ``jump" now often has a specific name: ``State reduction". One can consider this as a technical designation for the above ``jump". However, with the words ``State reduction" are also used the synonyms as ``Collapse of the state" or ``wave function collapse". 

The quantum particle during the measuring process is not a closed quantum system and hence it is not described by the Schr\"{o}dinger equation. The measurable quantum system is described by the density matrix given in Eq. (\ref{6}) and the evolution of quantum particle is connected with the full Hamiltonian of the combined quantum system QS+MD+ES presented in Eq. (\ref{8}). Hence, the measured quantum system can not be described by the wave function or ket-vector. In fact the measuring process can be described by the general scattering theory (see Appendix A) and hence the words as ``wave function collapse" are from a category of misleading and they have no physical sense.  For example, in the book of J. J. Sakurai \cite{Sak} and many others known books and monographs no such words and terms at all.  Actually the mathematical and physical foundation of quantum measurement theory has a long history and different approaches and concepts \cite{1a,2a,3a,4a,5a,7a}.

\section{Conclusions}

In this paper, we presented the axiomatic theory of quantum first-kind measurements. The measuring process is described with the ideal initial conditions (Postulates 1--4) and
non-ideal initial conditions (Postulate 5) which allow us to declare the theory of non-ideal measurements. 
It is shown in Sec. VI that the measurements of the momentum of particles are the most fundamental measurements
because appropriate probabilities and differential cross sections are the same
for ideal and non-ideal initial conditions. 
Hence, in this case no the survival effect and the theoretical computation
of the differential cross sections do not need any corrections connected to measuring procedure. 
The opposite situation arises for the measurements
in the position x-space which lead to
the survival effect described in Sec. VII. It is shown that the survival effect violates the Heisenberg uncertainty relation. Note that in this theory the bras and kets are distributions in the rigged Hilbert space which is an extension of
the Hilbert space. 

We describe the measuring device as a quantum system interacting with
an environment system. In this case, the non-ideal initial conditions are the consequence of the fact that the initial state of measuring device is metastable. We emphasize that in the axiomatic
theory of quantum measurements the probabilities arise in quantum mechanics as a result of measuring procedure and preparation of a mixed state of measurable QS. In the case of non-ideal initial conditions the probability distribution $P({\bf x},t_a)$ of particle in the x-space depends on the survival probability distribution $\tilde{{\rm P}}(t)$. Thus, in the general case, the wave function of particle does not allow to find the position probability distribution without the knowledge of the function $\tilde{{\rm P}}(t)$ which is a characteristic of the MD.  Nevertheless if the appropriate parameter $\epsilon$ is small the survival effect leads to small perturbation of the position probability distribution.

In the conclusion we note that the standard quantum mechanics describes the spectrums of observables, the quantum dynamics and scattaring of particles or, more generally, collision processes. The attempts to obtain a satisfactory solution of the quantum measurement problem based on the principles of quantum mechanics and statistical physics lead to different schemes and models of quantum measurement. The axiomatic formulation of quantum measurements presented in this paper shows that the satisfactory solution of the quantum measurement problem can be based on a rigorous approach using the rigged Hilbert space and appropriate set of Postulates describing the measurement process.

\section*{ACKNOWLEDGMENTS}

The author is grateful to Gerard Milburn for support of this work to present the quantum measurement theory in the axiomatic form. It was also supported in part
by the Physics Department of the University of Queensland.

\appendix

\section{SCATTERING WAVE PROPAGATOR}

In this Appendix, we introduce the the scattering wave propagator ${\cal U}_\nu^{in}(t)$ which is an important element of the scattering theory.
This approach yields a limiting process for computation of the functionals in a rigged Hilbert space. 
The Hamiltonian in the scattering theory has the form ${\rm H}={\rm H}_0+{\rm H}_{\rm I}$ where ${\rm H}_{\rm I}$ describes the interaction of the particles. 
The retarded solution of the Schr\"{o}dinger equation with the Hamiltonian ${\rm H}$ can be written in the following explicit form, 
\begin{equation}
|\Psi(t)\rangle=\nu\int_{-\infty}^0 e^{\nu t'}\exp\left(-\frac{i}{\hbar}{\rm H}(t-t')\right)|\Phi(t')\rangle dt',
\label{1a}
\end{equation}
where $|\Phi(t')\rangle=\exp(-it'{\rm H}_0/\hbar)|\Phi\rangle$. We show below that this equation leads to the Lippmann-Schwinger equation and hence it is a formal solution of the Schr\"{o}dinger equation with an appropriate boundary condition for scattering problem. It is assumed that this formal solution of the Schr\"{o}dinger equation can be used for calculation of the probabilities and 
differential cross sections with after transition to limit $\nu\rightarrow 0^{+}$ ( $\nu>0$) in the final step for evaluation of the appropriate functionals. More precise limiting process is presented in the end of this Appendix. Equation (\ref{1a}) has the form,
\begin{equation}
|\Psi(t)\rangle={\cal U}_\nu^{in}(t)|\Phi\rangle,
\label{2a}
\end{equation}
where ${\cal U}_\nu^{in}(t)$  is the scattering wave propagator given by
\begin{equation}
{\cal U}_\nu^{in}(t)=\nu\int_{-\infty}^0 dt' e^{\nu t'}{\rm U}(t-t'){\rm U}_0(t')={\rm U}(t)\Omega_\nu^{(+)},              
\label{3a}
\end{equation}
with ${\rm U}(t)=\exp(-it{\rm H}/\hbar)$ and ${\rm U}_0(t)=\exp(-it{\rm H}_0/\hbar)$. 
The wave operator $\Omega_\nu^{(+)}$ describing ingoing waves at $t<0$ is given by 
\begin{equation}
\Omega_\nu^{(+)}=\nu\int_{-\infty}^0 d\tau e^{\nu \tau}{\rm U}^{\dag}(\tau){\rm U}_0(\tau).
\label{4a}
\end{equation}
Another wave operator $\Omega_\nu^{(-)}$ describing outgoing waves at $t>0$ is  
\begin{equation}
\Omega_\nu^{(-)}=\nu\int_0^{+\infty} d\tau e^{-\nu \tau}{\rm U}^{\dag}(\tau){\rm U}_0(\tau).
\label{5a}
\end{equation}
The scattering wave propagator ${\cal U}_\nu^{out}(t)={\rm U}_0(t)\Omega_\nu^{(-)\dag}$ describing the outgoing waves is presented in Eq. (\ref{9}).
Note that the M\o ller wave operators in a scattering theory of particles are 
$\Omega_{+}={\rm lim}_{t\rightarrow -\infty}{\rm U}(-t){\rm U}_0(t)$ and 
$\Omega_{-}={\rm lim}_{t\rightarrow +\infty}{\rm U}(-t){\rm U}_0(t)$ where the limits of the operator sequences are defined in a sense of strong convergence. However, such definition have sense only for special class of the Hamiltonians. Such systems are known as the scattering systems. For the scattering systems, the operators $\Omega_\nu^{(\pm)}$ are equivalent (with appropriate limiting transition) to the M\o ller wave operators $\Omega_{\pm}$.
The wave operators $\Omega_\nu^{(\pm)}$ are isometric: 
\begin{equation} 
\Omega_\nu^{(+)\dag}\Omega_\nu^{(+)}={\bf 1},~~~\Omega_\nu^{(-)\dag}\Omega_\nu^{(-)}={\bf 1}.
\label{6a}
\end{equation}
In the general case, the operators $\Omega_\nu^{(\pm)}$ are not unitarian because the following relation is valid $\Omega_\nu^{(\pm)}\Omega_\nu^{(\pm)\dag}={\bf 1}-{\cal R}$ where ${\cal R}$ is the projection operator to subspace of bound states with interaction operator ${\rm H}_{\rm I}$. If no bound states then the operators $\Omega_\nu^{(\pm)}$ are unitarian.
The scattering operator $S$ is defined as
\begin{equation}
S=\Omega_\nu^{(-)\dag}\Omega_\nu^{(+)}.
\label{7a}
\end{equation}
The operator $S$ is unitarian: $SS^{\dag}=S^{\dag}S={\bf 1}$.  The unitary property of $S$ follows from Eq. (\ref{6a}) and the relation $\Omega_\nu^{(\pm)\dag}{\cal R}=0$.

We define the eigenstates $|\Phi_\lambda\rangle$ and eigenvalues ${\rm E}_\lambda$ of free Hamiltonian ${\rm H}_0$:
\begin{equation}
{\rm H}_0|\Phi_\lambda\rangle={\rm E}_\lambda|\Phi_\lambda\rangle.
\label{8a}
\end{equation}
Equation (\ref{2a}) at $t=0$ yields equation $|\Psi_\lambda(0)\rangle={\cal U}_\nu^{in}(0)|\Phi_\lambda\rangle$ where the state $|\Psi_\lambda(0)\rangle$ is given by
\begin{equation}
|\Psi_\lambda(0)\rangle=\int_{-\infty}^0 dt' \nu \exp(\nu t')\exp\left(\frac{i}{\hbar}({\rm H}-{\rm E}_\lambda) t'\right)|\Phi_\lambda\rangle.
\label{9a}
\end{equation}
This equation was introduced in the formal theory of scattering \cite{GMG}. Integration in Eq. (\ref{9a}) leads to equation,
\begin{equation}
|\Psi_\lambda(0)\rangle=\frac{i\hbar\nu}{{\rm E}_\lambda-{\rm H}+i\hbar\nu}|\Phi_\lambda\rangle,
\label{10a}
\end{equation}
where $|\Psi_\lambda(0)\rangle=\Omega_\nu^{(+)}|\Phi_\lambda\rangle\equiv|\Psi_\lambda^{(+)}\rangle$. Equation (\ref{10a}) also 
can be written in the form,
\begin{equation}
|\Psi_\lambda^{(+)}\rangle=|\Phi_\lambda\rangle+\frac{1}{{\rm E}_\lambda-{\rm H}+i\hbar\nu}{\rm H}_{\rm I}|\Phi_\lambda\rangle.
\label{11a}
\end{equation}
This equation with some algebraic transformations leads to the Lippmann-Schwinger equation \cite{LS},
\begin{equation}
|\Psi_\lambda^{(+)}\rangle=|\Phi_\lambda\rangle+\frac{1}{{\rm E}_\lambda-{\rm H}_0+i\hbar\nu}{\rm H}_{\rm I}|\Psi_\lambda^{(+)}\rangle.
\label{12a}
\end{equation}
We define the function $f_{\mu\lambda}(t)$ as
\begin{equation}
f_{\mu\lambda}(t)=\langle\Phi_\mu(t)|\Psi_\lambda(t)\rangle=\langle\Phi_\mu|\exp\left(\frac{i}{\hbar}({\rm E}_\mu-{\rm H}) t\right)    
|\Psi_\lambda\rangle,
\label{13a}
\end{equation}
where $|\Psi_\lambda\rangle=|\Psi_\lambda(0)\rangle$.
Equations (\ref{12a}) and (\ref{13a}) yield the relation,
\begin{equation}
f_{\mu\lambda}(0)=\delta_{\mu\lambda}+\frac{1}{{\rm E}_\lambda-{\rm E}_\mu+i\hbar\nu}{\rm T}_{\mu\lambda},
\label{14a}
\end{equation}
where ${\rm T}_{\mu\lambda}=\langle\Phi_\mu|{\rm H}_{\rm I}|\Psi_{\lambda}^{(+)}\rangle$.
It follows from Eqs. (\ref{12a}) and (\ref{13a})
the normalization condition for the amplitude $f_{\mu\lambda}(t)$ as
\begin{equation}
\sum_\mu |f_{\mu\lambda}(t)|^2=N_\lambda,
\label{15a}
\end{equation}
where $N_\lambda\equiv\langle\Psi_\lambda(t)|\Psi_\lambda(t)\rangle=\langle\Psi_\lambda^{(+)}|\Psi_\lambda^{(+)}\rangle$. Hence, the normalization parameter $N_\lambda$ is a constant. The probability that the system is in state $|\Phi_\mu\rangle$ at time $t$ is given by ${\rm w}_{\mu\lambda}(t)=|f_{\mu\lambda}(t)|^2/N_\lambda$. Equations (\ref{13a}), (\ref{14a}) and (\ref{15a}) lead (after appropriate computations) to normalization parameter $N_\lambda$ in the form,
\begin{equation}
N_\lambda=1+(\hbar\nu)^{-1}{\rm Im} {\rm T}_{\lambda\lambda}.
\label{16a}
\end{equation}
We have the relations $|\Phi_\lambda\rangle\propto \varepsilon^{3/2}$ and $|\Psi_{\lambda}^{(+)}\rangle\propto \varepsilon^{3/2}$ in the rigged Hilbert space (see Appendixes B and C). Hence ${\rm T}_{\lambda\lambda}\propto \varepsilon^3$ and the second term in Eq. (\ref{16a}) has the uncertainty of the type $\varepsilon^3/\nu$ when $\varepsilon,\nu\rightarrow 0$. To remove this uncertainty in the theory we have to assume that the parameters $\varepsilon$ and $\nu$ tend to zero as $\varepsilon^3/\nu\rightarrow 0^{+}$ when $\varepsilon,\nu\rightarrow 0^{+}$. Hence, the double limit in the functionals $F(\varepsilon,\nu,...)$ such as probabilities and cross sections should have the form,
\begin{equation}
\lim_{\nu\rightarrow 0^{+}} \lim_{\varepsilon\rightarrow 0^{+}} F(\varepsilon,\nu,...),
\label{17a}
\end{equation}
where the first limit is given at $\varepsilon\rightarrow 0^{+}$ and the second limit is given at $\nu\rightarrow 0^{+}$.
Thus, this double limit yields $N_\lambda=1$ and ${\rm w}_{\mu\lambda}(t)=|\langle\Phi_\mu(t)|\Psi_\lambda(t)\rangle|^2$. 

We note that usually it is used the Hilbert space
 in the scattering theory and the states $|\Phi_\lambda\rangle$ are normalized to unity in the cubic volume $L^3$. In this case, the second term in Eq. (\ref{16a}) is proportional to $\nu^{-1}L^{-3}$, and hence, it is uncertainty in the limit when $L\rightarrow \infty$ and 
$\nu\rightarrow 0$. To remove such uncertainty in the functionals, one should compute 
first the limit at $L\rightarrow \infty$ and second the limit at $\nu\rightarrow 0^{+}$.

\section{RIGGED HILBERT SPACE}

The rigged Hilbert space (Gelfand triplet or equipped Hilbert space) \cite{Gelf} is a triad of spaces ${\bf \Phi} \subset   {\cal H } \subset {\bf \Phi}^{\times}$ such that ${\cal H }$ is a Hilbert space, ${\bf \Phi}$ is a dense subspace of ${\cal H }$ and ${\bf \Phi}^{\times}$ is the space of antilinear functionals over ${\bf \Phi}$. The most significant examples of rigged Hilbert space (RHS) are those for which 
${\bf \Phi}$ is a nuclear space that is the abstract expression of the idea that ${\bf \Phi}$ consists of test functions and ${\bf \Phi}^{\times}$ consists of corresponding distributions. 
The space ${\bf \Phi}^{\times}$ is called the antidual space of ${\bf \Phi}$.
There is another RHS, ${\bf \Phi} \subset   {\cal H } \subset {\bf \Phi}'$, where ${\bf \Phi}'$ is called the dual space of ${\bf \Phi}$ and contains
the linear functionals over ${\bf \Phi}$. 
The basic reason why we need the space ${\bf \Phi}$ is that unbounded operators are not defined on the whole of ${\cal H }$ but only on dense subdomains of ${\cal H }$ that are not invariant under the action of the observables. Such non-invariance makes expectation values, uncertainties and commutation relations not well defined on the whole of ${\cal H }$. The space ${\bf \Phi}$ is the largest subspace of the Hilbert space on which such expectation values, uncertainties and commutation relations are well defined.
The bras and kets associated with the elements in the continuous spectrum of an observable belong, respectively, to ${\bf \Phi}'$ and ${\bf \Phi}^{\times}$ rather than to ${\cal H }$. Thus, the bras and kets are distributions in the rigged Hilbert space which is an extension of
the Hilbert space with distribution theory. Any selfadjoint operator possesses a complete system of generalized eigenfunctions in the RHS.
The same assertion is also correct for unitary operators acting on a rigged Hilbert space.

For an example, the generalized eigenvector $|\tilde{\phi}_\varepsilon;{\bf p}\rangle$ in a rigged Hilbert space with momentum ${\bf p}$ has the form, 
\begin{equation}
|\tilde{\phi}_\varepsilon;{\bf p}\rangle =\int \tilde{\phi}_\varepsilon({\bf p},{\bf p}')|{\bf p}'\rangle d{\bf p}',
\label{1b}
\end{equation}
where $\tilde{\phi}_\varepsilon({\bf p},{\bf p}')$ is a test function. The generalized eigenvector $|\tilde{\phi}_\varepsilon;{\bf p}\rangle$ is normalized which yields
\begin{equation}
\int |\langle {\bf p}'|\tilde{\phi}_\varepsilon;{\bf p}\rangle|^2d{\bf p}' =\int |\tilde{\phi}_\varepsilon({\bf p},{\bf p}')|^2d{\bf p}'=1.
\label{2b}
\end{equation}
In the particular case, the test function can be chosen as
\begin{equation}
\tilde{\phi}_\varepsilon({\bf p},{\bf p}')=\frac{1}{\pi^{3/4}\varepsilon^{3/2}}\exp\left(-\frac{1}{2\varepsilon^2}({\bf p}-{\bf p}')^2\right),
\label{3b}
\end{equation}
where $|\tilde{\phi}_\varepsilon({\bf p},{\bf p}')|^2=\delta_\varepsilon({\bf p}-{\bf p}')$ is a delta-sequence (with $\varepsilon\rightarrow 0$ for functionals). 
Equation (\ref{1b}) at $\varepsilon\rightarrow 0$  yields the asymptotic behaviour as
\begin{equation}
|\tilde{\phi}_\varepsilon;{\bf p}\rangle \sim 2^{3/2}\pi^{3/4}\varepsilon^{3/2}|{\bf p}\rangle. 
\label{4b}
\end{equation}
Thus we have the asymptotic relation $|\tilde{\phi}_\varepsilon;{\bf p}\rangle \propto \varepsilon^{3/2}|{\bf p}\rangle$ at $\varepsilon\rightarrow 0$.

\section{DISCRETE EIGENSTATES IN RHS}

It is introduced in Eq. (\ref{57}) the generalized states in momentum space which belong to the RHS.
These generalized states are orthonormal,
\begin{eqnarray}
\langle{\bf p}',\varepsilon|{\bf p},\varepsilon\rangle=\varepsilon^{-3}\int_{v_{{\bf p}'}} d{\bf p}_1\int_{v_{\bf p}} d{\bf p}_2 
 \langle{\bf p}_1|{\bf p}_2\rangle~~~
\nonumber\\ \noalign{\vskip3pt}  = \varepsilon^{-3}\delta_{ {\bf p}'{\bf p}}\int_{v_{\bf p}} d{\bf p}_1\int_{v_{\bf p}} d{\bf p}_2 
\delta({\bf p}_1-{\bf p}_2)=\delta_{ {\bf p}'{\bf p}}.
\label{1c}
\end{eqnarray}
Thus $\langle{\bf p}',\varepsilon|{\bf p},\varepsilon\rangle=\delta_{ {\bf p}'{\bf p}}$ where $\delta_{ {\bf p}'{\bf p}}$ is the Kronecker symbol.
We note that for $\varepsilon\rightarrow 0$ the following change is valid,
\begin{equation}
\varepsilon^{3}\sum_{\bf p}\rightarrow\int d{\bf p}. 
\label{2c}
\end{equation}
It follows from Eqs. (\ref{57}) and (\ref{2c}) the asymptotic relation at $\varepsilon\rightarrow 0$ as
\begin{eqnarray}
\sum_{{\bf p}}|{\bf p},\varepsilon\rangle\langle{\bf p},\varepsilon|=\varepsilon^{-6}\int d{\bf p}\int_{v_{{\bf p}}} d{\bf p}'\int_{v_{\bf p}} d{\bf p}'' 
|{\bf p}'\rangle \langle{\bf p}''|~~~
\nonumber\\ \noalign{\vskip3pt}  =\int |{\bf p}\rangle \langle{\bf p}|d{\bf p} ={\bf 1}.~~~~~~~~~~~~~~~~~~~~
 \label{3c}
\end{eqnarray}
Thus, the set of states $|{\bf p},\varepsilon\rangle$  is complete asymptotically in the RHS at $\varepsilon\rightarrow 0$.
We define the generalized momentum operator ${\bf P}$ in RHS as 
\begin{equation}
{\bf P}=\sum_{{\bf p}}{\bf p}|{\bf p},\varepsilon\rangle\langle{\bf p},\varepsilon|.
\label{4c}
\end{equation}
Hence, the states $|{\bf p},\varepsilon\rangle$ are the generalized eigenstates of the momentum operator: ${\bf P}|{\bf p},\varepsilon\rangle ={\bf p}|{\bf p},\varepsilon\rangle $. 

We can also introduce the generalized states in position space (x-space) which belong to RHS.
Thus, we introduce a discrete set of the coordinate vectors given by
\begin{equation}
{\bf x}=\varepsilon {\bf n}=\varepsilon(n_1,n_2,n_3),~~~~~~~~~n_k=0,\pm 1,\pm 2,...~,
\label{5c}
\end{equation}
where ${\bf n}$ is a vector with discrete components and $\varepsilon$ is an arbitrary positive parameter. We emphasise that the parameters $\varepsilon$ in Eqs. (\ref{54})  and  (\ref{5c})
 are different and independent. Moreover, they have different dimensions. We use, for simplicity, such notation when this does not lead to contradiction.
We can also define the cubic cell $v_{\bf x}$ with a discrete vector ${\bf x}$ as
\begin{equation}
\left\{{\bf x}'\in v_{\bf x}: \varepsilon\left(n_k-\frac{1}{2}\right)<{\rm x}_k'<\varepsilon\left(n_k+\frac{1}{2}\right)\right\},
\label{6c}
\end{equation}
where  $~~k=1,2,3$.
The volume of the cubic cell $v_{\bf x}$ is $\varepsilon^3$. The test function $\phi_\varepsilon({\bf x},{\bf x}')$ can be chosen as
\begin{equation}
\phi_\varepsilon({\bf x},{\bf x}')=\left\{\begin{array}{ll}
\varepsilon^{-3/2}  & \mbox{if ${\bf x}'\in v_{\bf x}$}\\
0 & \mbox{otherwise}
\end{array}
\right. 
\label{7c}
\end{equation}
where ${\bf x}$ is the discrete vector given in Eq. (\ref{5c}).
The generalized eigenstate in the rigged Hilbert space for the test function given in Eq. (\ref{7c}) is
\begin{equation}
|{\bf x},\varepsilon\rangle =\int \phi_\varepsilon({\bf x},{\bf x}')|{\bf x}'\rangle d{\bf x}'=\varepsilon^{-3/2}\int_{v_{\bf x}} |{\bf x}'\rangle d{\bf x}'. 
\label{8c}
\end{equation}

The states $|{\bf x},\varepsilon\rangle $ are orthonormal,
\begin{equation}
\langle{\bf x}',\varepsilon|{\bf x},\varepsilon\rangle=\delta_{ {\bf x}'{\bf x}}.
\label{9c}
\end{equation}
Here $\delta_{ {\bf x}'{\bf x}}$ is the Kronecker symbol. The set of states $|{\bf x},\varepsilon\rangle$ is also complete asymptotically in the RHS at $\varepsilon\rightarrow 0$,  
\begin{equation}
\sum_{{\bf x}}|{\bf x},\varepsilon\rangle\langle{\bf x},\varepsilon|=\int |{\bf x}\rangle \langle{\bf x}|d{\bf x} ={\bf 1}.
\label{10c}
\end{equation}

The proof that the set $|{\bf x},\varepsilon\rangle $ is orthonormal and complete asymptotically is the same as it is given in Eqs. (\ref{1c}) and (\ref{3c}) for the generalised eigenstates in a momentum space.
We define the generalized position operator ${\bf X}$ in the RHS as 
\begin{equation}
{\bf X}=\sum_{{\bf x}}{\bf x}|{\bf x},\varepsilon\rangle\langle{\bf x},\varepsilon|.
\label{11c}
\end{equation}
Hence, the states $|{\bf x},\varepsilon\rangle$ are the generalized eigenstates of the position operator: ${\bf X}|{\bf x},\varepsilon\rangle ={\bf x}|{\bf x},\varepsilon\rangle $.
The decomposition of the density matrix $\rho(t)$ of quantum system in a rigged Hilbert space at $\varepsilon\rightarrow 0$ has the form, 
\begin{equation}
\rho(t)=\sum_{{\bf x}'}\sum_{{\bf x}''}  \langle {\bf x}',\varepsilon|\rho(t)|{\bf x}'',\varepsilon\rangle |{\bf x}',\varepsilon\rangle \langle{\bf x}'',\varepsilon|.
\label{12c}
\end{equation}

\section{POSITION UNCERTAINTY}

In this Appendix, we use the next notations: $P(x)\equiv P(x,t_a)$ and $\tilde{P}(x)\equiv \tilde{P}(x,t_a)$ where the distributions
$P(x,t_a)$ and $\tilde{P}(x,t_a)$ are given by Eqs. (\ref{108})  and (\ref{109}) respectively. The moments $\langle x^n\rangle$ and
$\langle x^n\rangle_{0}$ are given by
\begin{equation}
\langle x^n\rangle=\int_{-\infty}^{+\infty}P(x)x^ndx,~~~\langle x^n\rangle_{0}=\int_{-\infty}^{x_0}P(x)x^ndx,
\label{1d}
\end{equation}
where $x_0=-a^2/2l$ and $n=0,1,2,...$~. In this case, we have the following relation,
\begin{equation}
\langle x^n\rangle_{0}=\frac{(-a)^n}{2\sqrt{\pi}}\left[\Gamma\left(\frac{n+1}{2},\sigma\right)-\sigma^{-1/2}\Gamma\left(\frac{n+2}{2},\sigma\right)\right],
\label{2d}
\end{equation}
where $\sigma=a^2/4l^2$, and the generalized Gamma function $\Gamma(\lambda,\sigma)$ is given by
\begin{equation}
\Gamma(\lambda,\sigma)=\int_{\sigma}^{+\infty}t^{\lambda-1}e^{-t}dt.
\label{3d}
\end{equation}
The function $\Gamma(\lambda,\sigma)$ has the asymptotics at $\sigma\rightarrow +\infty$ as
\begin{equation}
\Gamma(\lambda,\sigma)\sim \sigma^{\lambda-1}e^{-\sigma}\left[1+\frac{(\lambda-1)}{\sigma}+\frac{(\lambda-1)(\lambda-2)}{\sigma^2}+...    \right].
\label{4d}
\end{equation}
Equations (\ref{2d})  and (\ref{4d}) lead to the following asymptotic relation,
\begin{equation}
\langle x^n\rangle_{0}\sim \frac{(-1)^{n+1}a^n}{4\sqrt{\pi}}\left(\frac{a}{2l}\right)^{n-3}\exp\left(-\frac{a^2}{4l^2}\right)+...~.
\label{5d}
\end{equation}
The moments with renormalized (positive) probability distribution $\tilde{P}(x)$ have the form,
\begin{equation}
\langle x^n\rangle_{r}=\int_{-\infty}^{+\infty}\tilde{P}(x)x^ndx=\frac{\langle x^n\rangle}{Q}-\frac{\langle x^n\rangle_{0}}{Q},
\label{6d}
\end{equation}
where the normalization constant is $Q=\int_{x_0}^{+\infty}P(x)dx$. Thus, we have the following asymptotic relations at 
$\epsilon_0\rightarrow 0$ ($\epsilon_0\equiv 2l^2/a^2$):
\begin{equation}
\langle x\rangle_r\sim \frac{l}{Q}\left[1-\frac{l}{\sqrt{\pi}a}\exp\left(-\frac{a^2}{4l^2}\right)+...\right]~,
\label{7d}
\end{equation}
\begin{equation}
\langle x^2\rangle_r\sim \frac{a^2}{2Q}\left[1+\frac{l}{\sqrt{\pi}a}\exp\left(-\frac{a^2}{4l^2}\right)+...\right]~.
\label{8d}
\end{equation}
The asymptotic relation for the normalization constant is given by
\begin{equation}
Q\sim 1+\frac{2}{\sqrt{\pi}}\left(\frac{l}{a}\right)^3 \exp\left(-\frac{a^2}{4l^2}\right)+...~,
\label{9d}
\end{equation}
where the parameter $l$ is defined in Eq. (\ref{119}).
It follows from Eq. (\ref{9d}) that $Q-1\ll1$ when $\epsilon_0\ll 1$. Equations (\ref{7d}), (\ref{8d}) and (\ref{9d}) yield relations
$\langle x\rangle_r=l$ and $\langle x^2\rangle_r=a^2/2$ for $\epsilon_0\ll 1$. Moreover, we have equations $\langle x\rangle_r=\langle x\rangle$ and $\langle x^2\rangle_r=\langle x^2\rangle$ for $\epsilon_0\ll 1$. Hence, the distribution $P(x)$ and the probability distribution $\tilde{P}(x)$ both lead to the same average position $\langle x\rangle_r=\langle x\rangle=l$ and position uncertainty 
relation $\Delta x=\sqrt{a^2/2-l^2}$ when the condition $\epsilon_0\ll 1$ is satisfied.


\end{document}